# Emission of nitrogen-vacancy centers in diamond shaped by topological photonic waveguide modes

Raman Kumar[1], Chandan[1,*], Gabriel I. López Morales[1], Richard Monge[1], Anton Vakulenko[1,2], Svetlana Kiriushechkina[1,2], Alexander B. Khanikaev[1,2,3], Johannes Flick[1,3,4], and Carlos A. Meriles[1,3,†]

As the ability to integrate single photon emitters into photonic architectures improves, so does the need to characterize and understand their interaction. Here, we use a scanning diamond nanocrystal to investigate the interplay between the emission of room-temperature nitrogen-vacancy (NV) centers and a proximal topological waveguide. In our experiments, NVs serve as local, spectrally broad light sources which we exploit to characterize the waveguide bandwidth as well as the correspondence between light injection site and directionality of wave propagation. Further, we find that near-field coupling to the waveguide influences the spectral shape and ellipticity of the NV photoluminescence, hence allowing us to reveal nanostructured light fields with a spatial resolution defined by the nanoparticle size. Our results expand on the sensing modalities afforded by color centers, and portend novel opportunities in the development of on-chip, quantum optics devices leveraging topological photonics to best manipulate and readout single-photon emitters.

The interaction between solid-state color centers and engineered photonic structures has become a cornerstone in efforts to control light–matter coupling at the quantum level[1]. Platforms such as the nitrogen-vacancy (NV) center in diamond[2] offer interrelated optical and spin transitions, making them attractive for integration with photonic architectures that enhance emission, mediate interactions, and enable efficient readout[3-7]. While significant progress has been made in resonantly coupling color centers to conventional photonic environments — such as dielectric cavities[8], waveguides[9], and photonic crystal resonators[3] — comparatively less attention has been given to their interaction with topological photonic structures[10]. These systems offer unique advantages, including robustness to disorder[11-14], chiral edge transport[15,16], and novel optical mode confinement[17,18] not found in traditional photonic designs. Exploring how quantum emitters behave in such environments opens the door to new, mostly uncharted regimes of control and functionality[19], recently illustrated through the introduction of topological lasers[20].

Examining the impact of such structures on the color center response, however, requires precise spatial control of the emitter position and flexible optical access. Circumventing the limitations of systems engineered to host color centers at fixed positions[21,22], prior work has employed a range of approaches, including optical trapping of single nanodiamonds[23] and grafting of emitter-hosting nanoparticles to the tip of a near-field scanning optical microscope[24-27] (NSOM). In addition to controlling emitter placement, these techniques allow for direct access to the electromagnetic near field, which has been exploited, e.g., to visualize optical mode profiles and the spatial distribution of the local density of states (LDOS)[23,25] through changes in the fluorescence intensity or excited state lifetime. The use of NV centers in many of these preceding demonstrations is not coincidental given the rapidly growing set of broader scanning sensing applications they are enabling[28-34].

Here, we study the interaction between a small, room-temperature ensemble of nano-diamond hosted nitrogen-vacancy (NV) centers and the waveguide forming at the interface between topologically trivial and non-trivial photonic crystal lattices. We alternatively exploit color centers as sub-wavelength photon-sources or as near-field probes, a dual role that simultaneously allows us to characterize the spectral response of the waveguide and examine its impact on the photoluminescence (PL) of the emitter, stimulate light propagation across the photonic structure with preferential directionality, and sense the local chirality of the near field through the effect it has on the NV PL ellipticity.

**NV optical spectroscopy near a topological metasurface**

In our experiments we investigate the waveguide forming at the interface between two hexagonal photonic crystal lattices produced in a silicon-on-insulator wafer via electron lithography and plasma etching (Fig. 1a). On one side of the interface, the triangular pits forming the hexagon are shifted inwards to yield a "shrunken" honeycomb lattice featuring trivial topology; the converse is true for the other, "expanded" side of the crystal where the altered ratio between intracell and intercell coupling strengths results in a non-zero Chern number[35,36]. Both the shrunken and expanded lattices feature a direct bandgap at the Γ point, whereas the interface supports two counter-propagating edge modes robust against backscattering (Fig. 1b).

[1]Department of Physics, CUNY- The City College of New York, New York, NY 10031, USA. [2]Department of Electrical Engineering, CUNY-The City College of New York, New York, NY 10031, USA. [3]CUNY-Graduate Center, New York, NY 10016, USA. [4]Center for Computational Quantum Physics, Flatiron Institute, New York, NY 10010, USA. *Present address: Department of Physics BMS College of Engineering Bengaluru, Bengaluru, India- 560 019. †E-mail: cmeriles@ccny.cuny.edu

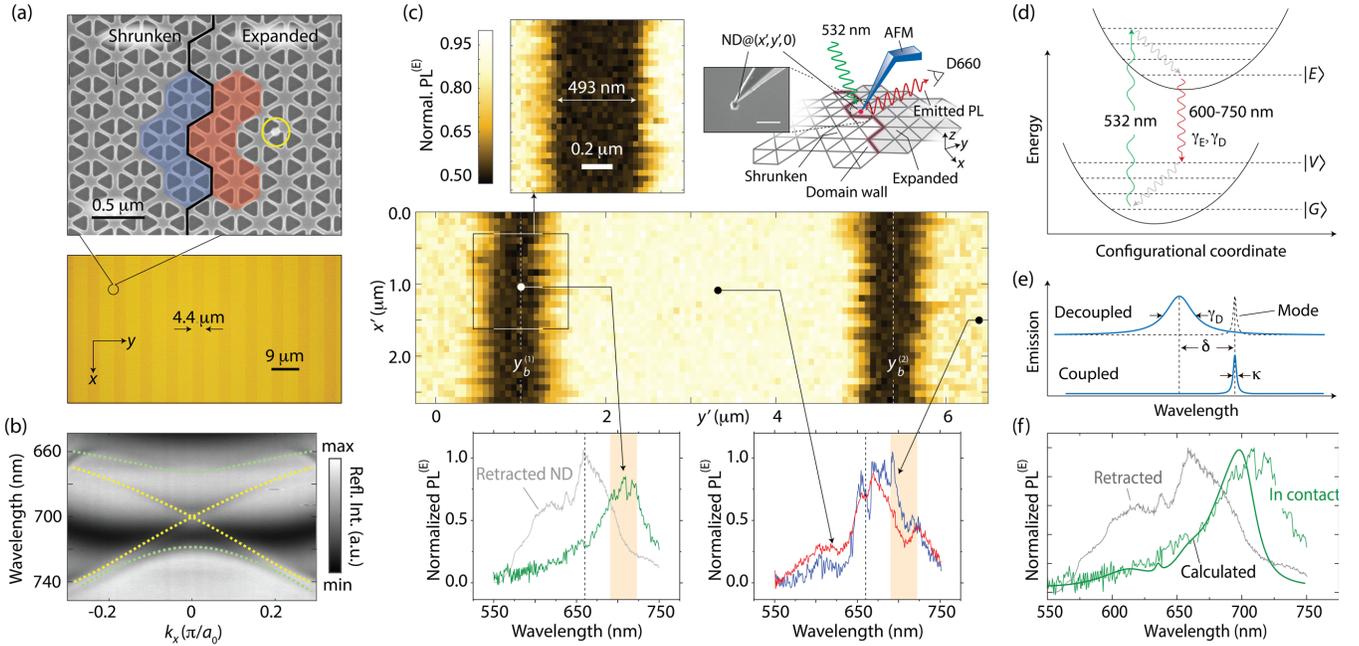

**Figure 1 | Scanning color center microscopy of a topological photonic waveguide.** (a) SEM and optical images (respectively, top and bottom) of the photonic crystal comprising alternating sections of expanded and shrunken hexagonal lattices (respectively, dark and bright orange stripes in the optical image). The yellow circle on the right-hand section singles out a typical nanodiamond (ND), overlaid on the surface. (b) Fourier-plane reflection spectrum on the domain wall showing dispersion of topological edge-states near the Γ point. Numerical results for a supercell eigenmode simulation are shown as dashed lines; green and yellow colors correspond to bulk and edge modes, respectively. (c) (Upper right) Experimental geometry. Coordinates $(x', y', z')$ indicate the ND position relative to the photonic crystal plane; D660 denotes photon detection in a 10-nm window centered at 660 nm. The upper insert is a scanning electron microscopy image of the tip and nanodiamond; the scale bar is 200 nm. (Main) Room-temperature PL imaging of a $4 \times 12$ μm$^2$ section of the photonic structure containing two domain walls at positions $y_b^{(1)}$ and $y_b^{(2)}$. The upper left insert is a zoomed image of the area in the square. (Lower inserts) Representative PL spectra in different sections of the photonic crystal; for reference, the gray trace shows the ND spectrum 100 μm above the crystal surface. The orange band indicates the photonic bandgap, and the dashed vertical line shows the detection wavelength. (d) Schematics of the NV emission dynamics. Relaxation from the excited state $|E\rangle$ into the ground state $|G\rangle$ typically relies on an intermediate, short-lived vibronic state $|V\rangle$. NVs undergo rapid pure dephasing with characteristic rate $\gamma_D$, much greater than the spontaneous decay rate $\gamma_E$. (e) Schematics of emission-tuning induced by pure dephasing. Coupling to a partly overlapping (but narrower) optical mode shifts the emission wavelength by $\delta$. (f) Calculated PL emission near the waveguide for a pure dephasing model that explicitly considers the phononic modes of both neutral and negatively charged NVs. Refl. Int: Reflected intensity.

We make use of an atomic force microscope (AFM) tip — hereafter referenced via coordinates $(x', y', z')$ — to pick up and manipulate an NV-hosting, 75-nm-diameter diamond nanoparticle previously drop-casted on a separate, flat silicon substrate; adhesion to the tip relies on van der Waals forces. We excite the NVs via a 532-nm laser and use a two-path confocal microscope to collect the PL directly emanating from the excitation point (i.e., emitted by the nanodiamond) or from a remote location upon scattering by the structure proximal to the tip (Supplementary Material (SM), Section I).

We first illustrate the nanodiamond (ND) response in Fig. 1c where we collect the PL emitted by the NV, PL$^{(E)}$, as we scan the particle across the photonic crystal; in this example, we limit photon detection to a 10-nm band centered at 660 nm. The resulting image clearly shows two trenches centered at the domain boundaries, here denoted $y_b^{(1)}$ and $y_b^{(2)}$. From comparing representative spectra across the boundaries and in bulk sections of the crystal (lower inserts in Fig. 1c), we conclude the PL change near $y_b^{(1,2)}$ arises from an overall shift of the NV emission to longer wavelengths. In all cases, we observe significant changes relative to the PL spectrum of a free-standing nanodiamond, hence exposing the non-trivial impact of the photonic structure on the NV emission.

We attribute these changes to phonon-assisted coupling of the NV emission to the waveguide, a process found in quantum emitters affected by fast pure dephasing rates $\gamma_D$[37,38]. This mechanism has been seen to shift the emission frequency of NV centers in a cavity to that of a narrower, far-detuned mode provided it partly overlaps with one or more NV vibronic sidebands, each homogeneously broadened due to fast phonon scattering[39] (Figs. 1d and 1e). A related process is at play near the waveguides, wherein 660-nm emission is partially suppressed and redirected to longer wavelengths, hence leading to dark trenches in the image. We qualitatively reproduce these observations assigning to the NV an effective phonon mode derived from density functional theory calculations (Fig. 1f, see also SM, Sections II through VII). We note this mechanism is fundamentally different from Purcell enhancement, which is negligible herein. In fact, data analysis shows that the integrated PL remains approximately constant in all areas of the photonic crystal, hence implying a redistribution, not a change, in the emitted power. Such reshaping of the emission spectrum can be exploited to introduce alternative forms of image contrast (SM, Section II).

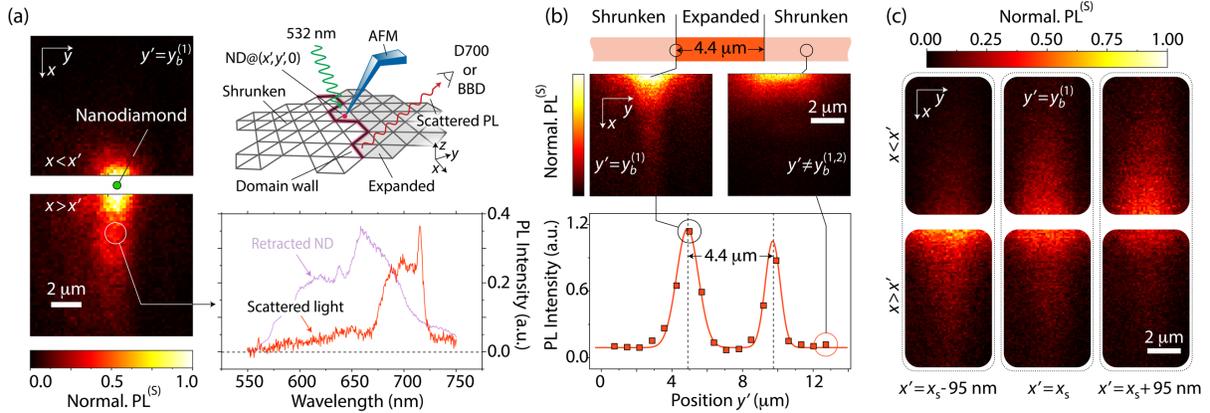

**Figure 2 | NV centers as a broadband light source.** (a) (Main) Confocal image of the NV PL upon scattering by the photonic crystal waveguide. In this example, the nanodiamond is in contact with the surface and has a fixed position $x'$ at the boundary $y' = y_b$; we observe directionality-selective propagation along the topological waveguide. (Lower right insert) Optical spectroscopy of the scattered PL. For reference, the faint, purple plot displays the ND emission spectrum 100 μm above the photonic crystal. (b) (Main) Scattered PL integrated across the region $x > x'$ as a function of the nanodiamond horizontal position $y'$. PL propagation centers around the photonic crystal boundaries; the upper inserts display representative scattered light images at one of the boundaries and away (left- and right-hand images, respectively). (c) Confocal images of scattered NV PL at different tip positions along a waveguide, i.e., $y' = y_b$. We observe a reversal of the propagation directionality as we displace the tip away from one of the symmetry points $x' = x_s$ (defined as a point where the NV PL is funneled equally in both directions). Except for the image in (a) — obtained via a broadband detector (BBD), 650 – 900 nm — all other images rely on photon detection after a 10-nm-bandpass filter centered at 700 nm (D700); the green laser power is 1 mw and all measurements are carried out under ambient conditions.

**The NV center as a local light source**

An appealing feature of color centers such as the NV is that the emitter can also be exploited as a local photon source over a broad spectral bandwidth. Leveraging the ability to collect light from a site away from that of laser excitation, we position the nanoparticle directly above one of the boundaries, i.e., $y' = y_b^{(1)}$, and map out PL$^{(S)}$, the NV PL upon scattering across the photonic crystal (Fig. 2a). We find a non-uniform pattern predominantly elongated along the direction of the waveguide (nearly aligned with the $x$-axis). The spectrum associated to the scattered PL centers at 700 nm and spans a ~50-nm bandwidth, much narrower than the emission of the free-standing ND. This range precisely matches that expected for the topological waveguide (Fig. 1b), and hence confirms the nature of the scattered luminescence.

We extend these observations in Fig. 2b where we measure the scattered PL amplitude as we displace the nanodiamond across two consecutive domain walls at positions $y_b^{(1)}$ and $y_b^{(2)}$. We find efficient photon waveguiding only within a ~1-μm-wide range centered at each wall; as shown in the scattered light maps above the main plot, there is negligible transmission when the nanodiamond hovers on the photonic crystal bulk.

Interestingly, the image in Fig. 2a shows that light propagation can be heavily one-directional, indicating preferential NV coupling to only one of the edge modes. The directionality of light propagation, however, depends sensitively on the point of photon injection and can be reversed by displacing the ND along the domain wall. We illustrate this response in Fig. 2c where we image the scattered PL on both sides of the nanoparticle at three different locations, $x_s$ and $x_s \pm 95$ nm, along $y_b$. We observe propagation along $x$ in the forward or backward direction depending on the sign of the displacement relative to the symmetry site, $x' = x_s$; this behavior repeats periodically as we displace the tip along the wall (SM, Sections IV through VII). These observations may at first be seen as a simple filtering effect, where the NV emission — unpolarized for a retracted nanodiamond, see SM,

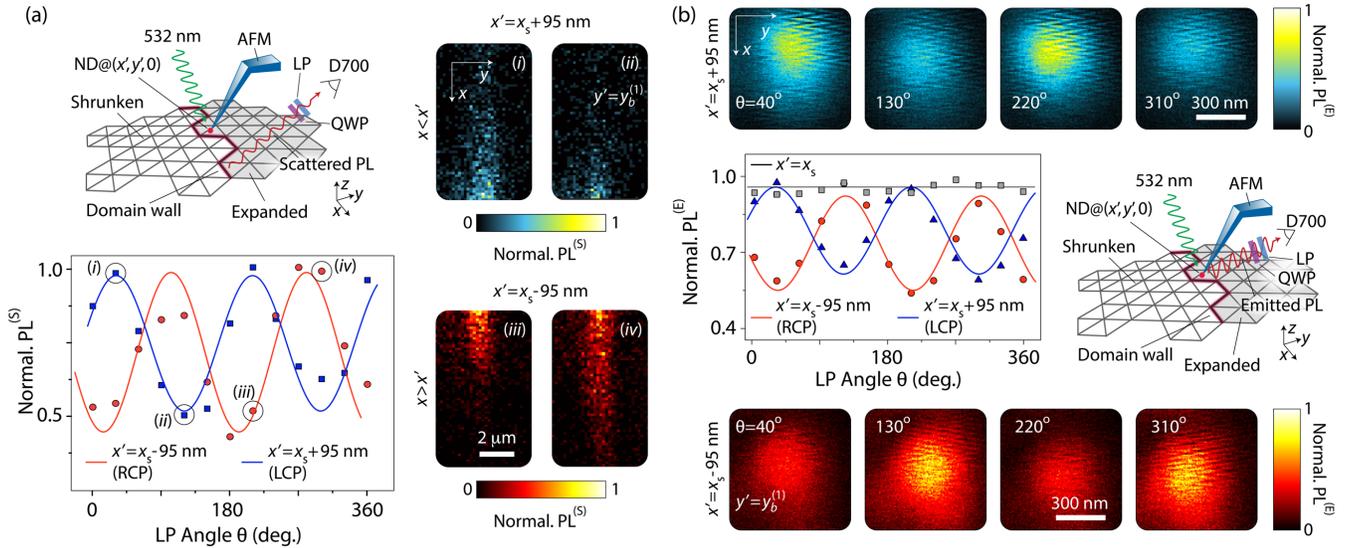

**Figure 3 | Correlated ellipticity between scattered and emitted PL.** (a) (Top and bottom left) Experimental geometry and measured scattered PL amplitude as a function of the linear polarizer angle $\theta$ for two different locations of the ND relative to the symmetry point $x' = x_s$; throughout these experiments, the photon collection point is 2 µm away from the ND along the waveguide. (Top and bottom right) Confocal imaging of the scattered PL for different settings of the polarizer angle $\theta$ ((*i*) through (*iv*) in the main plot). We find a correlation between the direction of light propagation and the preferential wave ellipticity. (b) Circular-polarization-selective confocal images of the nanodiamond (top and bottom inserts) and integrated NV photoluminescence dependence (middle plot) as a function of $\theta$ when displacing the nanodiamond ±65 nm relative to $x_s$. Unlike the case in (a), here the laser excitation (532 nm) and PL collection sites coincide during imaging. The NV emission's dominant ellipticity correlates with the local directionality (and circular polarization) of the scattered wave. QWP: Quarter wave plate. LP: Linear polarizer. D700: Detection with 10-nm bandpass filter centered at 700 nm.

Section IV — is selectively routed by the waveguide; we show below this is not the case.

**Polarization-sensitive imaging**

Topological modes forming at the boundaries between expanded and shrunken honeycomb lattices are known to be chiral, implying the direction of propagation correlates with light ellipticity[35,36]. We demonstrate this correspondence in Fig. 3a where we make detection selectively sensitive to right- or left-circularly polarized PL (RCP and LCP, respectively). Placing the nanodiamond before the symmetry point ($x' = x_s - 95$ nm) yields predominantly RCP scattered PL while the converse happens when the tip hovers past $x_s$ (e.g., $x' = x_s + 95$ nm). Remarkably, we find that this same ellipticity is imprinted in the luminescence collected directly from the NVs in the nanoparticle, as seen by plotting the PL response as a function of the detection polarizer angle for different positions of the nanodiamond (Fig. 3b). This behavior amounts to a generalization of preceding observations for a quantum dot emitter in a similar topological photonic crystal[16] (SM, Section V), and is reminiscent of that reported recently for NVs non-radiatively coupled to surface plasmon polaritons in a bullseye-like structure of circular nano-ridges with azimuthally varying width[40].

The high level of ellipticity in the NV emission — here expressed as a change in the NV brightness reaching up to 50% — again showcases the impact of the photonic structure on the NV response. We leverage this interaction in Fig. 4a where we monitor the NV fluorescence for two fixed configurations of the detection polarizer, respectively chosen to expose emitted RCP or LCP light as we displace the tip across the photonic crystal. Consistent with the observations in Fig. 3, imaging along the domain walls $y_b^{(1)}$ and $y_b^{(2)}$ shows alternating sections of predominantly right- and left-circularly polarized NV emission seamlessly alternating in space. This can be seen in Fig. 4b where we remove the quarter wave plate from our detection, thus resulting in the formation of nearly uniform dark trenches at the domain walls. The contrast virtually disappears in the absence of a polarizer (SM, Section II), which underscores the singular image formation mechanism at play, complementary to that observed in Fig. 1c.

Qualitatively, we associate fringe formation to preferential NV coupling with electromagnetic modes of a given helicity. This notion is supported by the one-on-one correspondence we find between fringe sites and the locations in the waveguide where scattered photons show preferred directionality (see SM, Section V). More rigorously, we can again rationalize our observations within the pure dephasing framework, with the NVs adjusting their emission wavelength *and* polarization to match the local electromagnetic modes they are coupling to most effectively (SM, Sections VI through IX). We caution, however, that a quantitative understanding of the observed polarization patterns will require additional work.

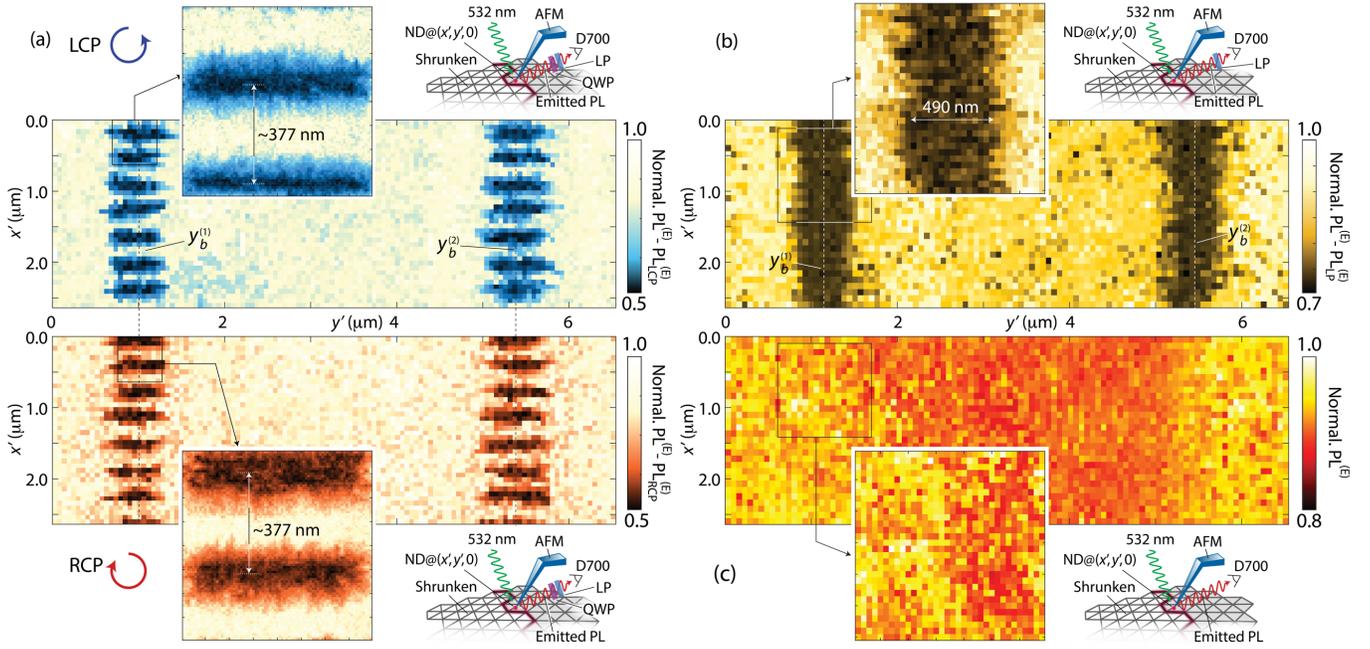

**Figure 4 | Polarization-sensitive imaging of the photonic crystal.** (a) Left- and right-circular-polarization-sensitive imaging of the topological crystal over an area spanning two consecutive domain walls; inserts are zoomed-in views of the area within black squares. We observe alternating PL drops along the boundaries, $y_b^{(1)}$ and $y_b^{(2)}$, between shrunken and expanded sections of the photonic crystal. (b) Same as in (a) but detecting after a linear polarizer. (c) Same as in (b) but after removing the polarizer. Compared to Fig. 1c, contrast is nearly lost close to the waveguide and enhanced between the bulk expanded and shrunken sections of the structure. In all schematics, QWP: Quarter wave plate; LP: Linear polarizer; D700: PL detection after a 10-nm-bandpass filter centered at 700 nm.

## Discussion

The singular light-matter coupling regime we find herein informs new approaches to integrating quantum emitters into photonic structures; these could be leveraged, e.g., to generate indistinguishable photons on chip[41], or to produce correlated photon states protected by topology[19]. Viewed as a sensing methodology, our findings complement preceding strategies to photonic structure imaging via quantum emitters, typically relying on changes of the PL intensity and/or lifetime. In the same vein, they add to the broader toolset in use today for near-field sensing[42], which besides NSOM[43-46], includes electron-based techniques, such as cathodoluminescence[47] and electron energy loss spectroscopy[48-51].

Additional studies will be needed to more accurately model the dynamics at work, both in terms of the phonon-assisted processes allowing the NV to alter its emission ellipticity, and the surprisingly large contrast we observe on the outgoing PL. By the same token, future work should investigate the role of ensembles versus individual NVs, as well as the impact of the photonic structure on NV spin pumping dynamics. Extensions to color centers other than the NV are also in order, particularly in systems exhibiting varying degrees of phonon-sideband emission; interesting illustrations include the SiV center in diamond[52], and color centers in two-dimensional materials[53]. Lastly, some of the processes observed here may find extension in the investigation of non-magnetic materials, for example, in the form of energy transfer imaging[54] with sub-diffraction resolution, or to discriminate enantiomers via the helicity of the emitted NV PL.

## Data availability

The data that support the findings of this study are available from the corresponding author upon reasonable request.

## Code availability

All source codes for data analysis and numerical modeling used in this study are available from the corresponding author upon reasonable request.

## Acknowledgments

We acknowledge helpful discussion with Vinod Menon and Toshu An. R.K. and C.A.M. acknowledge support by the U.S. Department of Energy, Office of Science, National Quantum Information Science Research Centers, Co-design Center for Quantum Advantage (C2QA) under contract number DE-SC0012704. S.K., A.V., and A.K. acknowledge support from the National Science Foundation (NSF) via grant NSF-2328993. J.F. acknowledges support from the NSF grant NSF-2216838. G.I.L.M. acknowledges NSF grant NSF-2208863. C. acknowledges NSF grant NSF-2203904; R.M. acknowledges support from NSF via grant NSF-2316693. All authors acknowledge access to the facilities and research infrastructure of the NSF CREST IDEALS, grant number NSF-2112550. The Flatiron Institute is a division of the Simons Foundation.

**Author contributions**

R.K., C., A.B.K., and C.A.M. conceived the experiments. C. led early experiments with drop casted nanoparticles showing directional propagation of PL; R.K. led the experiments in scanning probe geometry with technical assistance from R.M.; S.K., A.V., A.B.K., G.I.L.M., and J.F. carried out the modeling. All authors analyzed the data; C.A.M. wrote the manuscript with input from all authors.

**Competing interests**

The authors declare no competing interests.

**Correspondence**

Correspondence and requests for materials should be addressed to C.A.M.

# Emission of nitrogen-vacancy centers in diamond shaped by topological photonic waveguide modes

Raman Kumar[1], Chandan[1], Gabriel I. López Morales[1], Richard Monge[1], Anton Vakulenko[1,2], Svetlana Kiriushechkina[1,2], Alexander B. Khanikaev[1,2,3], Johannes Flick[1,3,4], and Carlos A. Meriles

[1]Department. of Physics, CUNY-City College of New York, New York, NY 10031, USA.
[2]Department of Electrical Engineering, CUNY-City College of New York, New York, NY 10031, USA.
[3]CUNY-The Graduate Center, New York, NY 10016, USA.
[4]Center for Computational Quantum Physics, Flatiron Institute, New York, NY 10010, USA.
[*]Present address: Department of Physics BMS College of Engineering Bengaluru, India- 560 019.

[†]E-mail: cmeriles@ccny.cuny.edu

# I. Experimental

*Confocal/atomic force microscope*

Our setup integrates a custom-built dual-galvo confocal microscope and an atomic force microscope (AFM). As shown in Fig. S1, the optical microscope has two paths, P1 and P2. Path P1 parks a continuous-wave (CW) green laser (532 nm) on the nanodiamond attached to the AFM tip using a combination of galvo mirror (GM1), some regular mirrors (RM), a dichroic mirror (DM, cut-on wavelength of 570 nm), and a two-lens 4f system. A 100× objective (Mitutoyo Plan Apo, 0.7 Numerical Aperture) focuses the green laser onto the nanodiamond. We collect the resulting fluorescence through path P2, which uses a similar 4f system, regular mirror (RM), a galvo (GM2), as well as a single photon counting module (Excelitas SPCM-AQRH). We reconstruct optical or fluorescence images by scanning GM2. A combination of quarter-wave plate (QWP) and a linear polarizer (LP) in path P2 let us characterize the polarization of emitted and scattered photoluminescence (PL), respectively denoted as $PL^{(E)}$ and $PL^{(S)}$. We precede the APD with a band-pass filter centered at different wavelengths to select the spectral region of interest as described in the main text.

The AFM is a modified Agilent system, which we use in conjunction with self-sensing and self-actuating Akiyama probes (Nanosensors); it operates in tapping-mode with a frequency of about 40 KHz. The bottom inset in Fig. 1 displays a schematic of the AFM control system: We use coarse positioners to align the tip with the optical objective while piezoelectric stages allow us to scan-move the sample relative to the tip with nanometer precision. The tip approaches the sample surface using a vertical stepper positioner. Upon interaction with the sample, the self-actuating Akiyama probe undergoes a change in oscillating frequency producing an error signal, which we then detect after amplification (NanoAndMore TFSC). A tuning fork sensor controller (NanoAndMore USA) provides the AFM feedback by passing the detected error signal to the AFM breakout box; the tip is constantly maintained by monitoring the error signal. Fig. S2a shows an image of the confocal microscope with the objective and AFM portion. Red arrows indicate the position of piezo scanners, sample, and tip position, as well as the preamplifier. Also visible in the figure is the green laser focused on the Akiyama probe.

Throughout our experiments, we use diamond nanoparticles from Adamas Nanotechnologies (NDNV120nm Hi) with an average size of 100 nm and an NV concentration of 3 ppm. We decorate the apex of the AFM tip with a nanodiamond by repeatedly scanning over a target particle; van der Waals forces from the combination of suspension residue on the particle and the native oxide on the silicon surface subsequently hold the particle in place as we bring it in contact with the sample surface[1]. Seen in Fig. S2b is an optical microscope image of the probe in contact with the photonic crystal under study. The SEM image of Fig. S2d displays the ~75-nm-diameter particle used in all experiments across the main text and supplementary material. We refer the reader to Section IX for additional information on nanoparticle/tip assembly and imaging reproducibility.

*The photonic crystal*

As discussed in-depth elsewhere[2,3], the photonic structure we exploited in our experiments closely resembles a quantum spin Hall system in condensed matter, where topological protections stem from time-reversal symmetry. The band structure in the range of interest can be modeled via a Dirac equation in which the effective mass — positive (negative) for compressed (expanded) regions — follows the spacings between triangles in a hexagonal cluster (Fig. S2e). As discussed in the main text (see also Figs. 1a and 1b), counterpropagating modes exist close to the wall separating the topological and trivial regions of the crystal, where the effective mass changes sign.

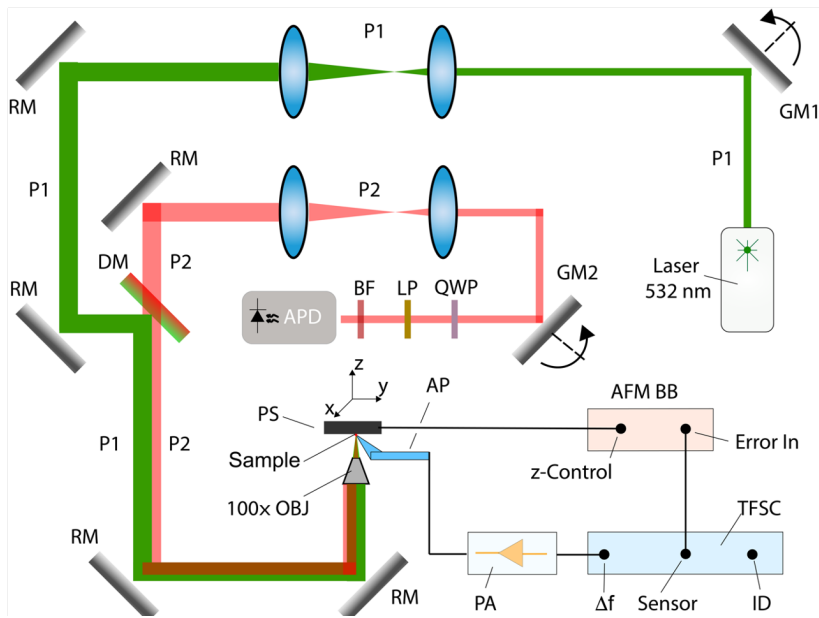

**Figure S1: Schematics of integrated dual-galvo confocal microscope and atomic force microscope.** P1: Path 1. P2: Path 2. GM1,2: Galvo mirrors; RM: Regular mirrors. DM: Dichroic mirror. QWP: Quarter wave plate. LP: Linear polarizer. BF: Bandpass filter. AFM: Atomic force microscope. APD: Avalanche photodetector. OBJ: 100× objective. TFSC: Tuning fork sensor controller. ID: Internal drive. Δf: Frequency change. AP: Akiyama probe. AFM BB: Atomic force microscope breakout box. PA: Pre-amplifier.

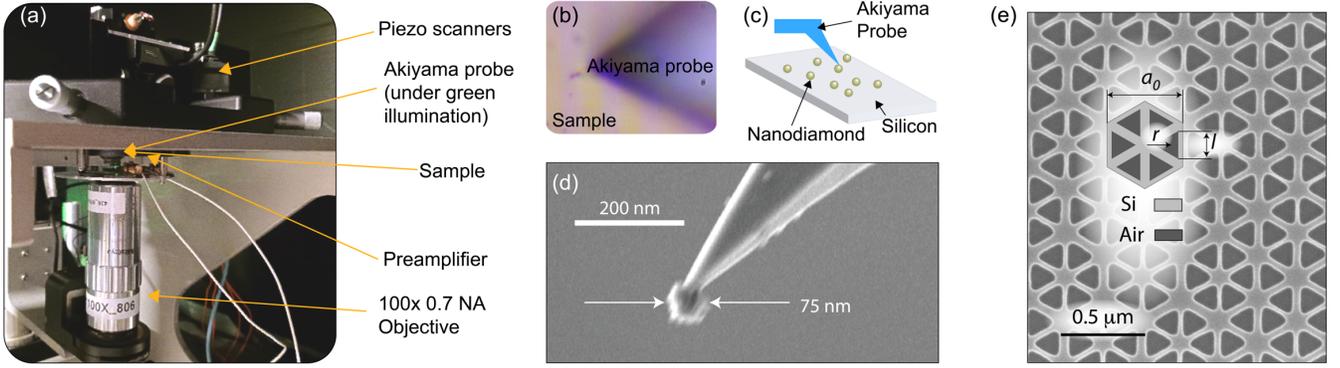

**Figure S2:** (a) For our experiments, we integrate a home-made tuning fork AFM (top) and a confocal microscope (bottom). (b) Optical microscope image of Akiyama probe in contact with the topological metasurface. (c) To pick up a nanodiamond on the tip, we simply scan the chosen particle multiple times until the AFM image vanishes. Subsequent optical and scanning electron microscopy (SEM) characterization confirm attachment to the tip. (d) SEM image of the tip; the nanodiamond diameter is 75 nm (reproduced from Fig. 1c in the main text). (e) SEM image of the photonic crystal and schematic. The structure we used has $a_0 = 450$ nm and a depth $h = 70$ nm; the radius separating the center of a shrunken (expanded) hexagon from the center of the surrounding triangles is $r_{sh} = 146$ nm ($r_{ex} = 161$ nm) and each equilateral triangle has a side length $l = 186$ nm.

Our fabrication protocol follows that already introduced in prior work[4]. Briefly, the photonic crystal structure was made out of a silicon-on-insulator (SOI) substrate comprising 70 nm of silicon on a 2-μm-thick buried oxide layer; patterning was carried out with the help of electron beam lithography (Elionix ELS-G100). We implemented the following procedure: We first spin-coated the substrate with a 150-nm-thick e-beam resist (ZEP520A-7), and then baked for 4 mins at 180 °C. To mitigate space charge field formation, we spin-coated the resist with a 50-nm-thick layer of an anti-charging agent (DisCharge H20x2). After e-beam exposure, we rinsed with DI water to remove the anti-charging agent, and developed the resist in *n*-amyl acetate at 0 °C for approximately 30 sec. We then exposed the sample to an inductively coupled plasma (Oxford PlasmaPro) and vertically etched the pattern to a depth of 70 nm. The etching recipe is based on C4F8/SF6 gases and achieves an etching rate of about 2.5 nm s$^{-1}$ at 5°C table temperature. Lastly, we removed any resist residues by immersing the sample in N-methylpyrrolidone solution heated to 60 °C. Figure S2e shows an image of the final structure along with a schematic of a unit hexagon (see caption for numerical values of all parameters).

## II. Spectral response of near-field-coupled NVs

We find that proximity of the diamond nanoparticle to the photonic structure causes significant changes in the NV emission spectrum, whose overall shape strongly depends on the particle location. We capture this dependence in Fig. S3a where we display representative NV PL spectra when the nanodiamond sits on one of the domain walls (upper green trace) or in bulk expanded and shrunken sections of the photonic crystal (red and blue traces, respectively). Comparison with the spectrum obtained after retracting the nanodiamond from the surface (black trace in Fig. S3a) shows the fluorescence emission preferentially moves to longer wavelengths. This red shift is most noticeable near the topological waveguide leading to peak emission in a ~30-nm-wide band around 710 nm, largely overlapping with the calculated crystal bandgap; it also approximately matches the wavelength range corresponding to the spectrum recorded after propagation in (and scattering off) the waveguide (Fig. 2a in the main text, see also Fig. S10 below). Spectral changes are also observable when the particle hovers on bulk sections of the crystal though the spectrum more closely resembles that of the retracted nanodiamond when over the expanded domain (compare red and black traces in Fig. S3a).

The relation between the shape of the NV emission spectrum and the location of the nanodiamond host provides opportunities for alternate forms of contrast. We demonstrate the idea in Fig. S3b where we compare scanning images of the same section of the photonic structure after limiting photon collection to 10-nm windows of variable central wavelength. In particular, we find that detection centered 700 nm virtually suppresses the contrast between the waveguide area and the shrunken domain (compare images (*ii*) and (*iii*) in Fig. S3b). Alternatively, detection at 720 nm reverses the contrast rendering the waveguide section brighter than the rest (images (*ii*) and (*iv*) in Fig. S3b). Note we can still identify the domain walls at shorter wavelengths where much of the NV emission is suppressed (see image (*i*) in Fig. S3b), though at the expense of a reduced contrast (and signal-to-noise ratio).

| Section of the photonic structure | Integrated area of emission spectrum (a.u.) |
|---|---|
| Near waveguide | 32.7±2 |
| Bulk crystal (expanded) | 34.9±2 |
| Bulk crystal (shrunken) | 34.1±2 |

**Table S1:** Comparison of integrated NV emission intensity for different positions of the nanodiamond host.

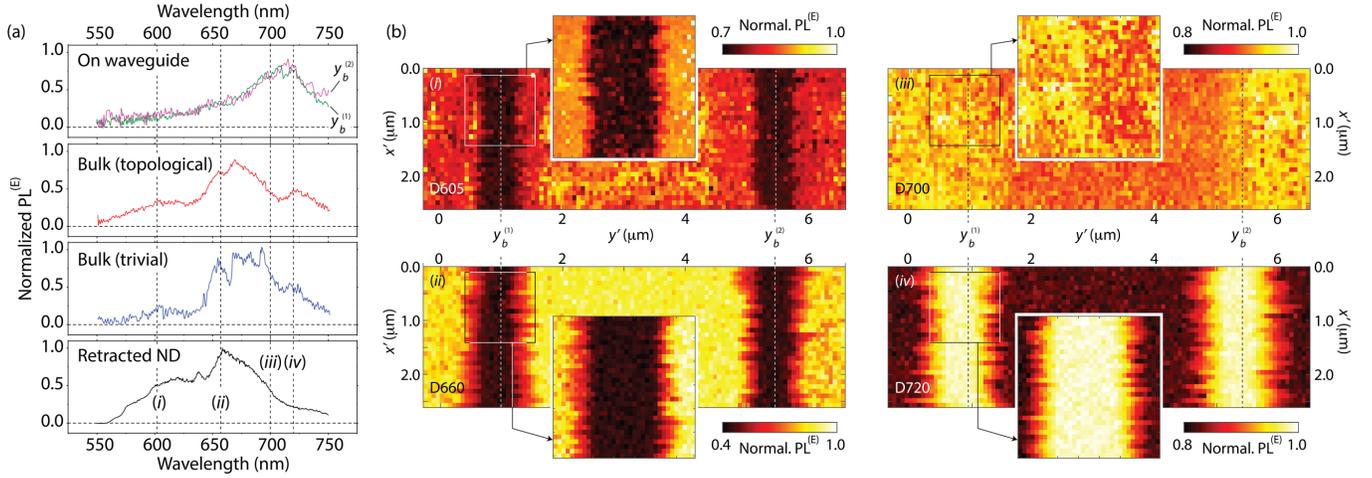

**Figure S3 | Spectral-contrast-based imaging of the photonic structure.** (a) Representative NV PL spectra under 532 nm excitation as we position the diamond nanoparticle in different sections of the photonic structure, namely, on the waveguide (green and purple traces), or on the bulk topological and trivial domains (red and blue traces, respectively); for reference, the black trace shows the PL spectrum after retracting the nanoparticle 100 µm from the crystal surface. The dashed vertical lines (and accompanying numeric labels) indicate the center wavelengths corresponding to the spectra in (b). Except the purple trace, all other spectra are reproduced from Fig. 1 in the main text. (b) Scanning imaging of the emitted NV PL in a section of the photonic crystal spanning two domain boundaries, $y_b^{(1)}$ and $y_b^{(2)}$. In each case, we use a 10-nm bandpass filter centered at a variable wavelength (denoted as D$\lambda$nm). Going from (*i*) to (*iv*), the center wavelengths are 605, 660, 700, and 720 nm, respectively. Note that (*ii*) has been reproduced from Fig. 1c in the main text. Normal.: Normalized to the maximum value.

Comparison of the integrated fluorescence in different sections of the photonic structure indicates that the overall change in the NV fluorescence intensity, if present, is small (all intensities coincide within 10%, see Table S1). Notice that comparing the integrated emission count rate near the photonic structure to that obtained after retracting the tip is unwarranted as light reflection from the surface can have a non-trivial impact on the collection efficiency.

The observations above suggest that despite the spectral reshaping we observe, the NV excited state lifetime remains roughly unchanged at different locations. We confirm this suspicion in Fig. S4, where we probe the temporal dependence of the NV spontaneous emission upon pulsed optical excitation at 532 nm. Overall, we observe inverse decay rates of order 20 ns, consistent with prior observations in nanodiamond[5]. Recording the NV response at different sites on the photonic structure, and comparing the result against that observed for a retracted nanodiamond, we find that changes in the NV excited state lifetime take place selectively in the vicinity of the waveguide. This change, however, is small (if at all present) again supporting the notion of a weak coupling to the waveguide and a pure-dephasing-driven process (as opposed to a Purcell-driven process). Indeed, in the limit where the relaxation channels of the photonic structure act over a spectral window narrower than the NV PL spectrum (see Section VI below), one expects a modified Purcell enhancement factor $\mathcal{F} \sim g^2/\gamma\gamma^*$ where $g$ is the coupling constant, $\gamma$ denotes the emitter's intrinsic linewidth, and $\gamma^*$ represents the homogeneously broadened linewidth. In the pure-dephasing-driven limit, $\mathcal{F}$ becomes negligible as $g \ll \gamma^*$. This is precisely the case reported by Albrecht et al. for NV centers coupled to a Fabry-Perot cavity[6], and is consistent with the observations herein.

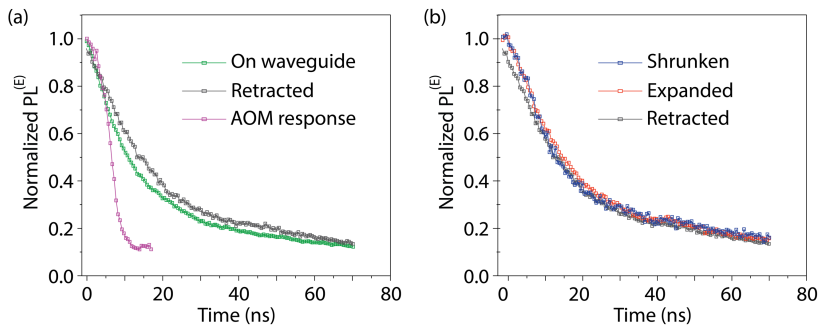

**Figure S4 | Site-selective lifetime measurements.** (a) NV PL intensity as a function of time after 532 nm excitation on the waveguide and retracted 100 µm from the structure (green and black traces, respectively). For reference, we include the measured acousto-optic modulator (AOM) response curve (purple trace). (b) Same as in (a) but with the nanodiamond on the expanded and shrunken bulk sections of the photonic crystal (red and blue traces, respectively).

### III. Does NV photochromism play a role?

NV centers are known to exist is different charge states, most typically negative and neutral, each of which features distinct fluorescence properties. Specifically, both NV forms are known to emit over a broad range in the visible and near infrared, but because the energy splitting between the ground and excited states of $NV^0$ is larger, its PL spectrum is blue-shifted relative to that of $NV^-$. Green illumination (e.g., 532 nm as used herein) excites both NV forms and can drive two-photon processes leading to $NV^-$ ionization and $NV^0$ recombination, a cyclic charge inter-conversion mechanism whose rate depends quadratically on laser intensity[7].

In bulk crystals of low to moderate nitrogen concentration (i.e., less than a few ppm), the NV charge state is metastable in the dark, with lifetimes exceeding several weeks[8]. This is not the case in doped systems where the preferred charge state depends on the crystal's Fermi level; NVs tend to be negatively charged in systems with high content of nitrogen since the latter acts like a donor. The diamond nanoparticles we used — designed to host the highest possible concentration of NVs — should fall in this category, except that surface-induced band bending significantly alters the relative band alignment and hence changes the equilibrium NV charge state[9]. In particular, shallow NVs tend to be neutral as surface traps effectively capture all donor electrons[10]. This process has proven thus far nearly inevitable in nanodiamond where surface oxidation — one of the known antidotes — is difficult to attain efficiently due to the multi-facet, irregular nature of the crystals.

To determine the charge state composition of the NVs in the diamond nanoparticle we used, we resort to known nanodiamond emission spectra[11] for $NV^0$ and $NV^-$, and fit a linear combination of the two to the PL spectrum observed far from the photonic crystal surface (Fig. S5a). Despite the moderate agreement, we conclude that a good fraction of these nanoparticle-hosted NVs are in the neutral charge state (($45 \pm 10$)%), a fact already apparent from the relatively large emission at wavelengths shorter than the $NV^-$ zero phonon line (637 nm). Note that due to the presence of a dichroic filter at 570 nm in our confocal microscope, a portion of the $NV^0$ spectrum is partially suppressed including its zero-phonon line. Therefore, we conclude that the evaluated fraction is a lower bound on the actual population of $NV^0$.

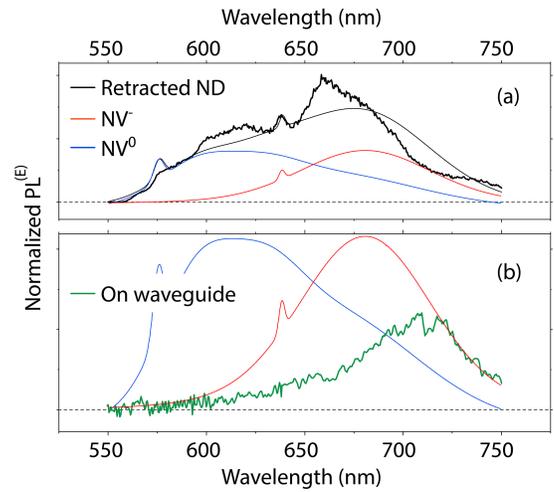

**Figure S5 | Comparison with charge-state-selective NV spectra.** (a) Far from the photonic structure, the NV PL spectrum can be qualitatively reproduced by a linear combination of $NV^0$- and $NV^-$-selective spectra (blue and red traces, respectively[11]). We estimate the $NV^0$ concentration at ~45%. (b) Comparison between the $NV^-$ PL spectrum and the measured ND spectrum near the waveguide.

We have seen above (Section II) that proximity to the photonic crystal tends to red shift the emission spectrum, hence raising the question as to whether the PL changes that we observe stem from a varying equilibrium charge state. Such scenario could conceivably result, for example, from electrostatic effects created by residues introduced during surface patterning of the silicon structure. To address this question, we carried out extensive comparisons between spectra corresponding to charge-state-specific NV spectra and our previous observations. Figure S5b shows one illustration corresponding to the case when the nanodiamond sits proximal to the waveguide: At first sight, the modified PL spectrum does seem to capture some of the features expected for a transformation from $NV^0$ to $NV^-$, most prominently the slow decay at longer wavelengths. A closer inspection, however, reveals this change also entails a significant depletion of photon emission below 700 nm, very much contrary to that observed for negatively charged NVs. Further, we find vanishing PL emission near the $NV^-$ zero phonon line at 637 nm, all of which strongly suggests the spectral features we observe do not stem from changes in the equilibrium NV charge state. This conclusion is also consistent with the observed dependence of the PL polarization on the nanoparticle location along the waveguide (see Figs. 3 and 4 in the main text as well as Sections VI and VII below), difficult to reconcile with a change in the NV charge state.

## IV. Polarized NV photoluminescence emission

This section provides additional information on the effect of the photonic structure over the polarization of the NV emission, which we condense in Fig. S6. Throughout these experiments, we restrict photon collection to a 10-nm window centered at 700 nm, where coupling of the emitters to the waveguide was found to be optimal (see Figs. 1a and 3b in the main text). We start in Fig. S6a where we retract the nanodiamond 100 μm from the photonic crystal surface to first determine the NV response in the absence of near-field effects. PL measurements after a linear polarizer show a marginal dependence on the polarizer's angle, even when we change the linear polarization of the excitation beam (532 nm) by 90 degrees. This latter result is not unanticipated, as rapid vibrational relaxation within the excited state manifold after off-resonant (i.e., green) illumination should erase any dependence on the excitation history. Further, room-temperature averaging of the NV excited states is expected to yield equal amounts of right- and left-circularly polarized PL photons, as well as photons linearly polarized along orthogonal axes[12,13]. Contrasting observations in individual NVs[14], the lack of polarization in the emission is exacerbated here by the large number of NVs probed (approximately 100 oriented along all four possible crystallographic directions) as well as the large strain/local electric fields present in nanodiamond (locally altering the orientation of the transition dipoles governing the emission of each individual NV).

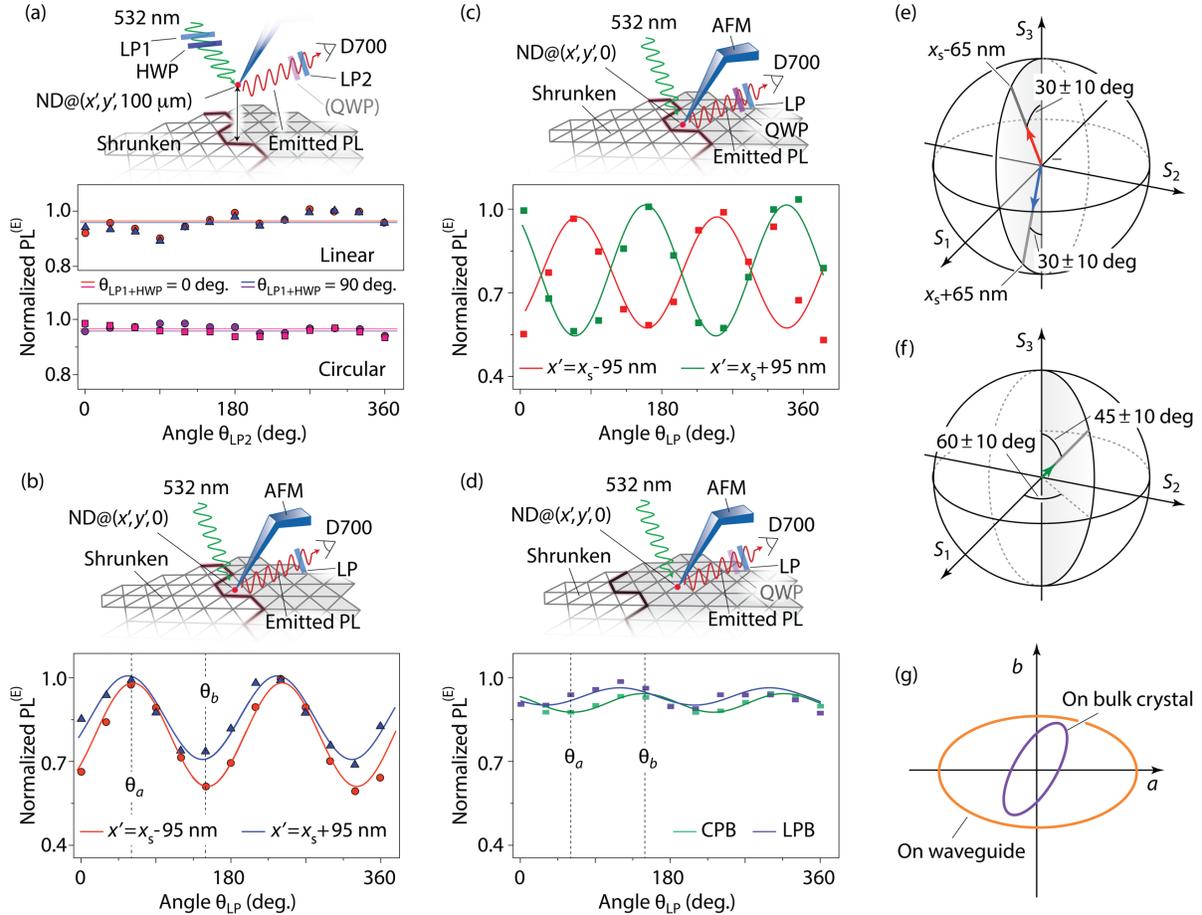

**Figure S6 | Linear polarization of the NV emission.** (a) We probe the degree of linear and circular polarization in the NV emission (top and bottom plots, respectively) for variable linear polarization of the excitation laser as the nanodiamond hovers far away from the surface ($z' \approx 100$ μm). The main plot shows the normalized NV PL as a function of $\theta_{LP2}$, the orientation of the detection polarizer, for two orthogonal polarizations of the excitation beam. In both instances, we observe no dependence on the orientation of the detection polarizer. (b) Same as in (b) but with the NV in contact with the photonic crystal at two points along a domain wall symmetrically displaced relative to one of the symmetry points, $x_s$. The laser output — which is linearly polarized — corresponds to the setting $\theta_{LP1+HWP} = 0$ deg. (c) Same as in (b) but in a configuration sensitive to circular polarization. The results match those presented in Fig. 3b in the main text. (d) Detection of circular and linear polarization (respectively, CPB and LPB) with the nanodiamond positioned on a bulk section of the photonic crystal. HWP: Half wave plate. LP: Linear polarizer. QWP: Quarter wave plate. D700: Detection within a 10 nm window centered at 700 nm. (e) Poincare sphere with $S_1$ through $S_3$ denoting the Stokes parameters. Red and blue arrows indicate the Stokes vector on the waveguide $\pm 65$ nm from a symmetry point $x_s$; the vector fractional size represents the degree of total polarization. (f) Same as in (e) but for the case where the nanodiamond sits on the bulk crystal. (g) Polarization ellipse as determined from (e) and (f) (orange and purple traces, respectively). The relative ellipse sizes — defined as the sum of the squares of the principal axes — have been chosen so as to match the ratio between the polarization intensities $I_p = (S_1^2 + S_2^2 + S_3^2)^{0.5}$ in each case. Orthogonal axes $a$ and $b$ correspond to the orientations of the linear polarizer as defined in (b).

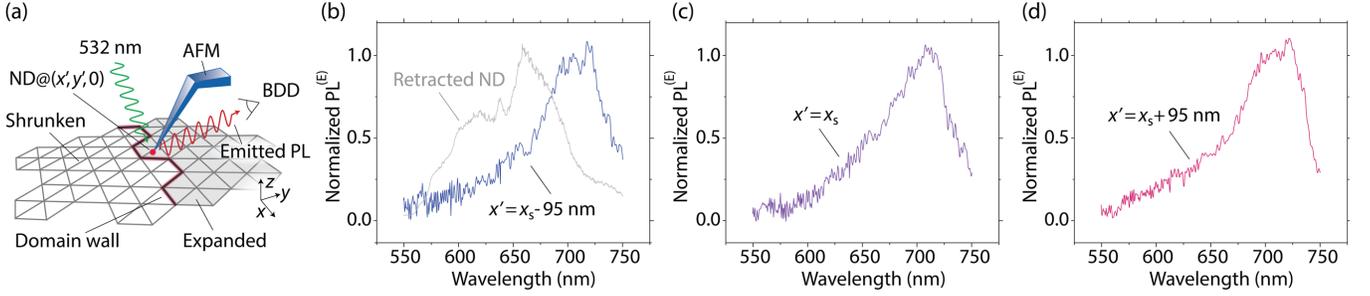

**Figure S7 | Impact of topological waveguide on the nanodiamond emission.** (a) We record the NV fluorescence spectrum as we displace the diamond nanoparticle along a domain boundary $y_b$. BBD: Broad-band detector. (b) Normalized NV fluorescence spectrum with the nanoparticle located 65 nm before $x_s$, the point of breakeven propagation. For reference, the faint trace shows the fluorescence spectrum with the tip retracted 100 μm from the photonic crystal surface. (c) Same as in (a) but with the nanocrystal located at position $x' = x_s$. (d) Same as in (c) but with the nanoparticle moved to $x' = x_s + 65$ nm. In (b) through (d), $y' = y_b$.

This behavior significantly changes when one brings the nanodiamond in proximity to the waveguide: As shown in Fig. S6b, we find a strong dependence on the orientation of the detection polarizer, thus revealing a high degree of linear polarization in the 700 nm emission. We observe only a secondary change as we displace the tip, here illustrated for sites along the waveguide where light injection from the nanodiamond yields photon propagation in opposite directions (see Fig. 2 in the main text). Using $\theta_a$ and $\theta_b = \theta_a + 90$ deg. to define a cartesian basis, we determine Stokes parameters $S_1$ and $S_2$ from the linear polarization contrast. For simplicity, we take an average over the waveguide, and find $S_1 \approx 0.43 \pm 0.05$ (by construction, $S_2 = 0.0$). We extend these observations in Fig. S6c where we intercalate a quarter wave plate to probe the circular polarization of the NV emission; from the measured contrast, we derive $S_3 \approx 0.70 \pm 0.07$. Using known relations for the Stokes vector components[15], we conclude the NV emission is elliptically polarized, the ratio between principal axes in the ellipse being 1.77. Denoting the fractional polarization intensity as $I_p$ and using the relation $\sum_{i=1}^{3} S_i^2 = I_p^2$, we conclude $I_p^{(w)} \approx 0.47 \pm 0.07$ where the superscript indicates a site in the photonic structure close to the domain wall. Note the dramatic change relative to the case when the nanodiamond has been retracted from the photonic structure surface (where $I_p^{(r)} \approx 0$), or when it has been displaced to the bulk crystal (where $I_p^{(b)} \approx 0.10 \pm 0.05$, see Figs. S6d through S6g).

Having investigated the interplay between the tip's position on the waveguide and the polarization of the outgoing PL, a natural question is whether the observed response is also accompanied by changes in the shape of the NV emission spectrum. We address this question in Fig. S7 where we compare the NV PL spectra for three different locations of the nanodiamond symmetrically distributed around one of the symmetry points (i.e., a point of breakeven propagation) along the waveguide. As shown in Figs. S7b through S7d, we find the spectral response — clearly different from that observed for the retracted tip, gray trace in Fig. S7b — remains largely unchanged for all three tip positions. This observation is consistent with the underlying symmetry of the density of states associated to mode sets with opposite pseudospin (see Fig. S4 in Section III), and thus the fact that, on average, NVs must couple equally to either mode set. The near-field nature of the coupling can locally break the symmetry in this interaction thus resulting in a varying PL polarization when the tip moves along the waveguide; the emission spectrum, however, remains unchanged as it can only reflect on the wavelength range of these edge modes and effective coupling amplitude to the emitter, both of which are (nearly) site-independent.

## V. Coupling to propagating waveguide modes

To demonstrate the near-field nature of the NV coupling to the waveguide, we reconstruct images of the NV fluorescence as it propagates (and scatters off) the photonic crystal structure for variable vertical heights of the diamond nanoparticle. We illustrate the system response in Fig. S8a for the case where the nanodiamond hovers above the waveguide at a site chosen to yield backward light propagation along the $x$-axis. For a given height, the mode amplitude diminishes with distance to the point of light injection as it radiates to the far field. The integrated scattered PL — shown in the main plot and extracted from the recorded images displayed in the upper inserts — decays quickly with the tip's vertical displacement $z'$ on a scale consistent with the evanescent nature of the mode. As indicated in Section I, technical constraints presently keep us from attaining fine control over $z'$, restricting our measurements to 100 nm steps. Practical limitations notwithstanding, we see virtually no light scattered off the waveguide above 200 nm, a height that here serves as an upper bound.

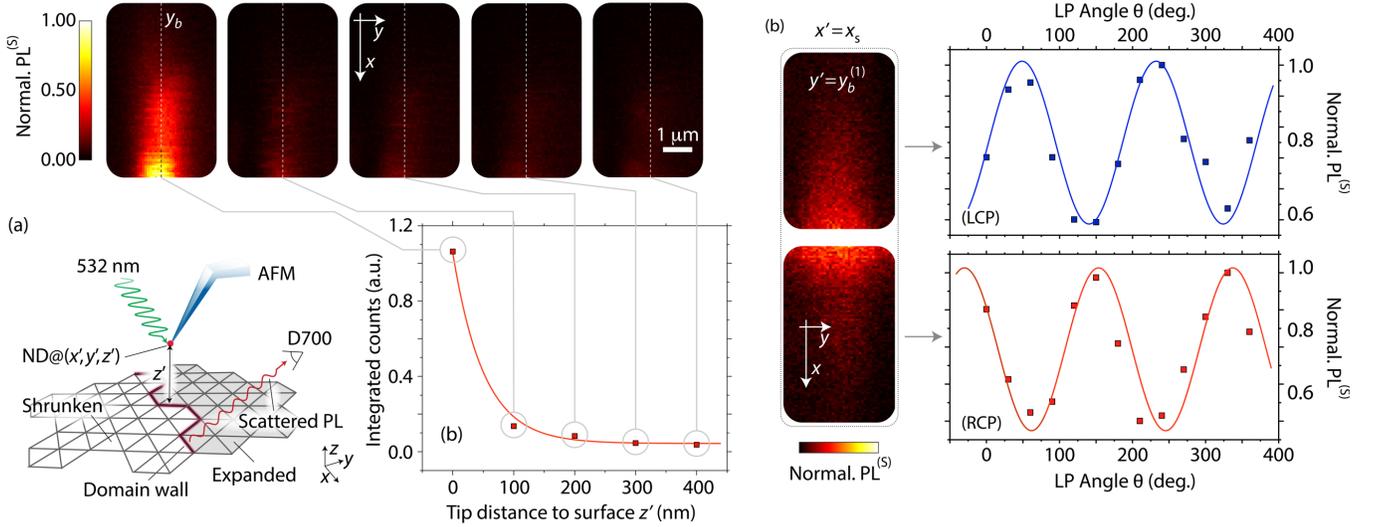

**Figure S8 | Dependence on distance to the waveguide.** (a) (Schematics) Confocal imaging of the NV PL scattered by the waveguide in the vicinity of a domain boundary at $y_b$. The nanodiamond (ND) sits at a fixed position on the waveguide but hovers at a variable height $z'$ relative to the surface; detection uses a 10-nm-bandpass filter centered at 700 nm (D700). (Main) Integrated PL scattered across the waveguide as a function of the nanodiamond vertical distance $z'$ to the surface; the upper inserts show the confocal images in each case. We observe a rapid decay of the waveguided emission within the first 100 nm, indicative of near-field coupling. The solid line is an exponential curve serving as a guide to the eye. (b) (Left) Scattered PL map for a nanodiamond proximal to the photonic crystal surface at a symmetry point $x' = x_s$ along the domain boundary $y' = y_b^{(1)}$ (reproduced from Fig. 2c in the main text). (Right) Scattered PL ellipticity for light propagating in the forward and backward directions along the waveguide (bottom and top plots, respectively). Photons propagating through the waveguide in one direction or the other separate according to their polarization.

Figure S8b introduces related observations for a nanodiamond closest to the structure at a symmetry point $x' = x_s$ along the waveguide, where the scattered PL map shows equal intensities of forward and backward propagating light (see Fig. 2c in the main text). As shown by the polarization-sensitive plots, we find that photons propagating in opposite directions preferentially feature opposite circular polarization, namely, right- and left-circular polarization for forward and backward propagation. These results are consistent with the complementary observations in Fig. 3, and allow us to conclude RCP (LCP) photons propagate in the forward (backward) direction regardless the point of injection (as expected for this class of topological waveguides[2,3], see also Sections VI and VII below). In other words, the asymmetric, position-dependent maps of scattered PL we measure in Fig. 2c of the main text serve as a proxy for the preferential polarization of photons emanating directly from the nanodiamond.

We explicitly demonstrate the interplay between emission ellipticity, position along the waveguide, and helicity of the excited mode in Fig. S9 where we compare side-by-side results obtained by collecting either the scattered or the emitted NV PL as the tip moves along the waveguide. Specifically, in Fig. S9a we measure the PL scattered off the waveguide 2 μm ahead of the light injection point, and observe periodic maxima corresponding to the points yielding forward light propagation. On the other hand, Fig. S9b displays an RCP-selective image of the same section of the waveguide; we find that the separation between the dark fringes — corresponding to areas where the NV PL is preferentially right-circularly polarized — precisely matches the points of forward light propagation across the waveguide. Note that because scattered light at these same sites is also right-circularly polarized, we find that helical modes in the waveguide imprint the ellipticity of the NV PL. Therefore, these results show how selective near-field coupling renders the quantum emitter a local reporter on the nanoscale structure of the electro-magnetic modes supported by the underlying structure.

It is interesting to compare the results above to prior observations in a similar topological photonic crystal[16], where light from a quantum dot (with emission wavelength within the photonic waveguide bandwidth) gets channeled into the waveguide along a direction dependent on its circular polarization. Unlike the present case, coupling to the structure in that study does not seem to alter the quantum response, i.e., the emitter yields an equal number of right- and left-circularly polarized photons. The reason for the difference is not presently clear but one possibility could be serendipity: As shown in Fig. S8b above, an emitter located in one of the "symmetry" positions along the waveguide — $x_s$ in Figs. 2 and 3 of the main text — does yield an equal number of photons propagating in either direction (each one featuring a characteristic photon ellipticity); at these sites the emitted PL appears unpolarized as in the retracted case (see polarization plot in Fig. 3b of the main text and Fig. S6 above). This behavior corresponds to that reported in Ref. 16, but because the quantum dot used in that study occupies a fixed position, the more general phenomenology we report here may have remained undetected. Note that although Barik et al. use a strong magnetic field to spectrally separate right- and left-circularly polarized emission (consistent with the observations in Fig. S8b and Fig. 3 of the main text), the quantum dot PL would look unpolarized if one was to measure the PL intensity (see Fig. 3b in Ref. 16). In this light, our findings can be seen as a generalization made possible by the singular dynamics afforded by pure-dephasing-driven dynamics and the flexibility intrinsic of the scanning probe geometry.

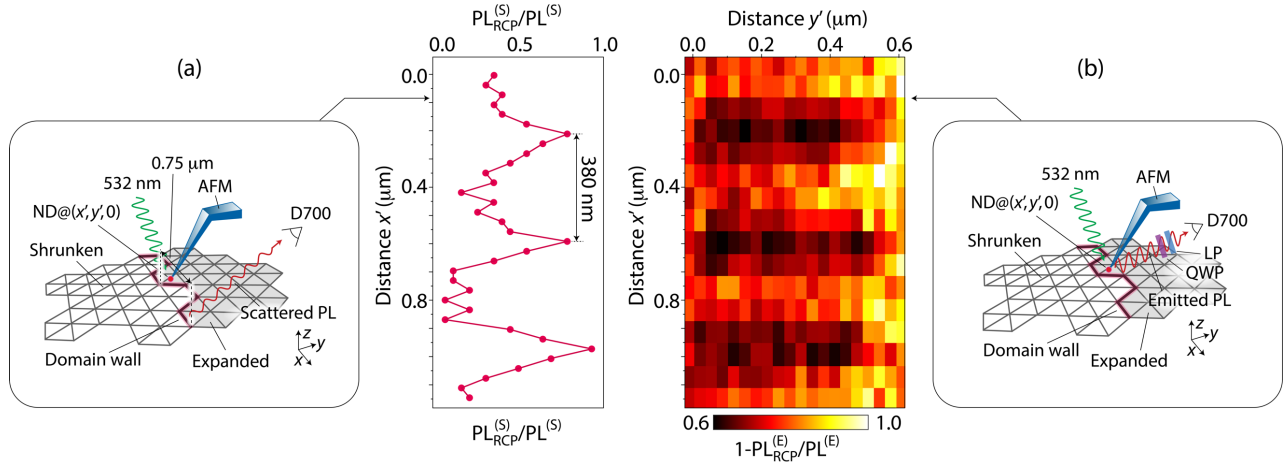

**Figure S9 | Correspondence between propagation directionality and emission ellipticity.** (a) We monitor the amplitude of the scattered photoluminescence at position $x' + 2$ μm as we change the tip position $x'$ along the domain boundary (i.e., $y' = y_b$). Intensity peaks in the plot correspond to instances of preferential propagation along $+x$. (b) Scanning probe image of an area along the boundary with detection polarizers configured to block $PL_{RCP}$. We observe good correlation between the image dips (associated to maximum RCP emission ellipticity) and the forward propagation peaks in (a). LP: Linear polarizer. QWP: Quarter wave plate. D700: Detector preceded by a 10-nm bandpass filter centered at 700 nm.

**VI. Rationalizing NV emission dynamics**

We posit that the observed changes in the NV center emission — both in terms of spectral shape and PL polarization — are a consequence of pure optical dephasing, fast for NVs due to phonon scattering at room temperature. In this regime, interaction with the photonic structure does not alter the emitter's lifetime (a feature we demonstrated in Section II) implying Purcell enhancement is negligible; rather, the system's emission shifts to match the resonance frequency of the coupled electro-magnetic mode. Theoretical[17,18] and experimental[19,20] studies have examined this phonon-assisted process for the case of individual quantum dots coupled to a cavity; more recent work zeroed in on diamond nanoparticles coupled to a tunable Fabry-Perot cavity, and showed that the emission from an individual NV can be tuned simply by altering the separation between the cavity mirrors[6]. In this latter case, the NV spontaneous emission was modeled with the help of seven vibronic levels whose brightness and lifetimes were chosen to match the NV zero-phonon-line and Stokes-shifted PL spectrum.

Compared with a cavity, the richer mode structure of the photonic crystal comparatively complicates our ability to formally capture the NV emission process. We find, however, clear signatures of pure-dephasing-driven dynamics by comparing the photonic-crystal-modified emission against the calculated energy density time average (EDTA) spectrum near the domain wall as well as the shrunken and extended bulk sections of the crystal (Fig. S10). Specifically, we find in all three cases that the PL emission spectrum of the near-field-coupled NVs shifts to the spectral range where the EDTA — here serving as a proxy for the local density of states — is higher. In particular, the NV spectra on both the expanded and shrunken sections of the photonic crystal feature local maxima near 720 nm; this behavior is in stark contrast with that observed for the retracted diamond nanoparticle, whose PL spectrum decays monotonically within that same frequency range (lower trace on the left side of Fig. S10). The correspondence is less apparent on the waveguide because the side maxima in the calculated EDTA seem to be absent in the measured spectrum (green trace in Fig. S10).

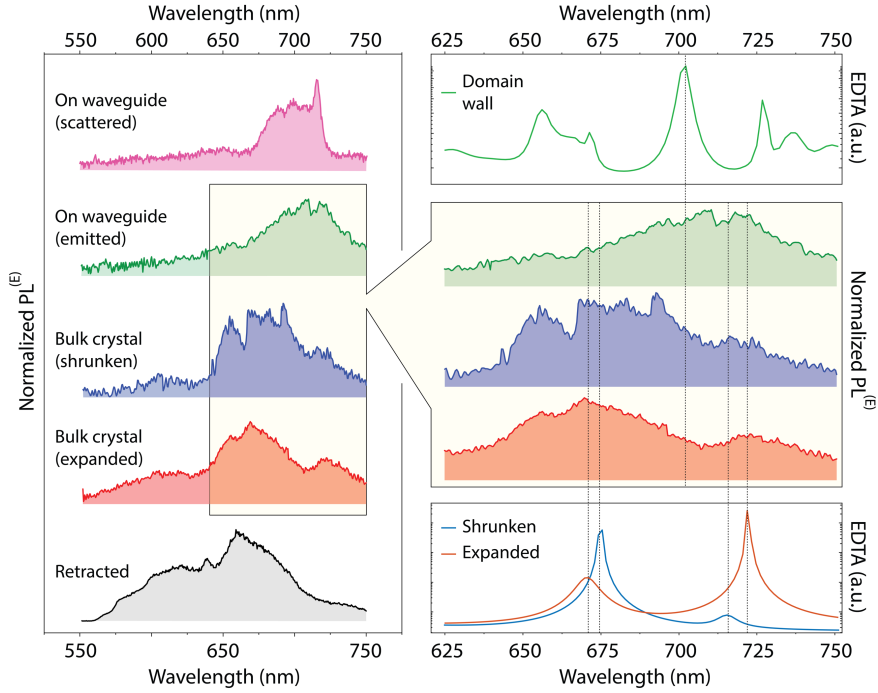

**Figure S10 | Spectral imprinting of photonic modes.** (Left) Measured spectra at different sites on the photonic structures (reproduced from Figs. 1 and 2 in the main text). (Right) Calculated energy density time average (EDTA) on the waveguide as well as the expanded and shrunken sections of the photonic crystal (top and bottom plots, respectively) vs zoomed PL spectra in the relevant spectral window. On the bulk section of the crystal, the local emission maximum near 720 nm suggests a response driven by the enhanced density of states in that wavelength range. On the waveguide, the NV PL emission approximately peaks in the region where the density of states is also maximum.

Note, however, these maxima are also absent in the PL spectrum scattered by the waveguide, which rather tends to match that collected from direct NV emission (compare pink and green traces on the left-hand side of Fig. S10).

To qualitatively capture the NV dynamics in the presence of the photonic waveguide, we use a simplified picture where photon emission derives from a decay of an excited state $|E\rangle$ into a vibronic state $|V_i\rangle$, which subsequently relaxes non-radiatively into the ground state $|G\rangle$ (Fig. S11a). Since the presence of a proximal diamond nanoparticle amounts to a defect in the photonic waveguide, the combined system can be effectively modeled as an "atom" coupled to a lossy "cavity"[21]. We therefore write the Hamiltonian as

$$H = \omega_{E,i}\sigma_+\sigma_- + \omega_W a^\dagger a + g(a^\dagger \sigma_- + a\sigma_+), \tag{S1}$$

where $\sigma_\pm$ represent the spin ladder operators, $a$ denotes the photon annihilation operator, and $g$ is the emitter coupling to the waveguide, here simplified to host a single mode of frequency $\omega_W$, and we express $H$ in units of $\hbar$. Note that in general, $\omega_{E,i}$ differs from the emitter's zero-phonon line (ZPL) by a frequency $\omega_i$ depending on the vibronic mode considered (Fig. S11a, see also below). Further, the Hamiltonian in Eq. (S1) does not account for photon polarization, a simplification assumed here for ease in presentation but revisited in the next section.

The Hamiltonian in Eq. (S1) can be similarly applied to NV[-] or NV[0] with the understanding that $\omega_{E,i}$ refers to charge-state-selective vibronic transitions. This point is important since the nanoparticle's PL spectrum suggests an appreciable population of NV[0] coexisting with NV[-] (see Fig. S5). Therefore, we consider these two charge states as independent emitters, and evaluate the waveguide effects on their excited-state dynamics separately as we do not anticipate the waveguide to impact the NV photoionization rates. In particular, we construct the set of vibronic states for NV[-] based on the effective phonon energy $\omega_P = 63.3$ meV, as derived from first-principles and in good agreement with experiment[22]. For the case of NV[0], we take $\omega_P = 45.0$ meV based on the observed phonon replicas[23]. Similarly, we assign to NV[-] the ZPL transition frequency $\omega_{ZPL} = 470.200$ THz and spontaneous relaxation rate $\gamma = 48$ MHz; the pure dephasing rate is $\gamma^* = 14.5$ THz and a vibronic relaxation rate $\gamma_P = 80$ THz, similar to those used in previous work[6]. For NV[0], $\omega_{ZPL} = 520.899$ THz, $\gamma = 35$ MHz[5], and $\gamma^* = 20.4$ THz (based on the best fit to the PL spectrum). In the same vein, we model the waveguide as a "cavity" featuring a single mode at $\omega_W = 424.8$ THz and a loss rate $\kappa = 4.6$ THz, which we derive phenomenologically. Lastly, we set the coupling constant at a value $g = 4$ GHz, corresponding to $0.3 - 0.4$ GHz per emitter for an ensemble of $N \approx 100 - 200$ NVs, approximately an order of magnitude lower than used before for a nanodiamond-hosted single NV in a tunable cavity[6]; note that since $g \ll \gamma^*, \gamma_P$, the system never experiences Purcell enhancement, consistent with experiment (see Section II).

To calculate the NV emission spectrum in the presence of the cavity, we assume that the master equation governing the evolution of the system's density matrix, $\rho$, has a Lindblad form with relaxation channels corresponding to spontaneous emission, photon loss, and pure dephasing (see below), all of which we code using the Python library

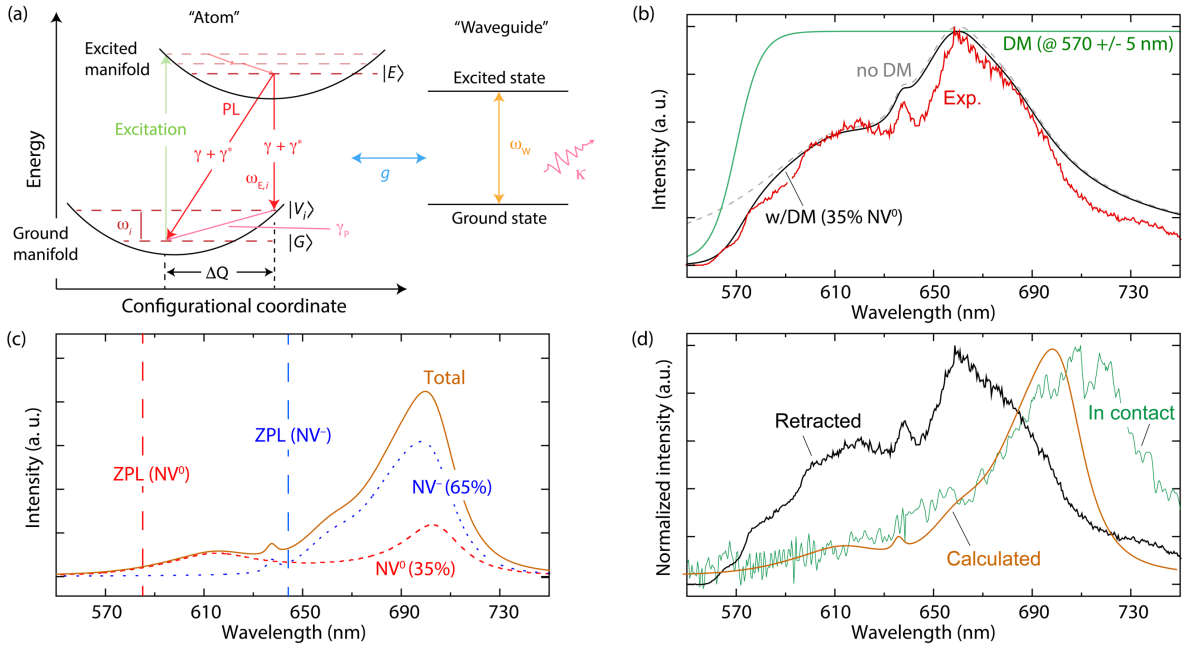

**Figure S11 | NV emission dynamics in the presence of a photonic waveguide.** (a) We model the NV emission as a process of spontaneous relaxation from the excited state manifold into a vibronic state $|V_i\rangle$. When in contact with the nanoparticle, the waveguide can be described as an effective cavity with resonance frequency $\omega_W$. (b) Experimental and modeled NV emission spectra (red and black traces, respectively) away from the waveguide. Here, we assume a 35% contribution from NV$^0$ and a dichroic mirror (DM) with a 570-nm cutoff (green trace). The dashed grey line is the simulated spectrum in the absence of a dichroic mirror. (c) Calculated NV emission in the presence of the photonic waveguide for the conditions in the black trace of (b). The blue and red traces respectively indicate the response from NV$^-$ and NV$^0$; the vertical dashed lines indicate their respective zero-phonon lines. (d) Comparison between the calculated NV PL spectrum as derived in (c) (black trace), and the NV emission when retracted from or in contact with the photonic waveguide (red and green traces, respectively).

QuTiP[24]. The emission spectrum follows from the quantum regression theorem and can be calculated via the two-operator, one-time correlation function[25], namely,

$$I_S(\omega) = \frac{\Gamma_S}{\pi} \int_{-\infty}^{\infty} \langle O^\dagger(\tau), O(0) \rangle e^{i\omega\tau} d\tau, \tag{S2}$$

where the subscript $S = E, W$ refers to the system under consideration, i.e., the emitter or the waveguide, $\Gamma_S$ indicates the relevant relaxation rate, and $O$ is the corresponding operator (e.g., $\kappa$ and $a$ for the waveguide, respectively). Following Ref. [6], we express the NV spectrum as stemming from contributions corresponding to spontaneous relaxation from the excited state to vibronic and the ground states (i.e., the ZPL transition). We use the measured NV emission in the absence of the waveguide to derive the relative weights $A_i$ of each relaxation channel; the PL spectrum takes therefore the form

$$I_E(\omega) = C_o \sum_i A_{0,i} I^{(0)}_{NV,i}(\omega) + C_- \sum_i A_{-,i} I^{(-)}_{NV,i}(\omega), \tag{S3}$$

where $C_o$ and $C_-$ are fitting parameters quantifying contributions from either NV charge state. Assuming the coupling is weak with respect to the detuning or to the emitter and waveguide mode's widths, the action of the photonic waveguide can be captured through the simple formula[17,18]

$$I_E^{(E-W)}(\omega) = I_E(\omega) \times I_W(\omega). \tag{S4}$$

|  | $A_0$ | $\gamma_0^*$ (THz) | $A_1$ | $\gamma_1^*$ (THz) | $A_2$ | $\gamma_2^*$ (THz) | $C_0$ |
|---|---|---|---|---|---|---|---|
| **NV$^0$** | 0.25 | 20.4 | 1.0 | 141.4 | 10.0 | 306.3 | 0.35 |
| **NV$^-$** | 1.0 | 14.5 | 16.0 | 219.9 | 1.0 | 86.4 | 0.65 |

**Table S2**: Numerical values of all parameters in Eq. (S3).

Figures S11b through S11d summarize our results: We can reproduce the measured NV emission spectrum reasonably well when considering two phonon replicas for both NV$^0$ and NV$^-$ as listed in Table S2. In the presence of the waveguide, we find that both charge state contributions undergo a substantial shift to longer wavelengths centered around the waveguide mode; Fig. S11d shows the end result, in qualitative agreement with our observations. In assessing these results, we emphasize our model must be seen solely as a conceptual framework able to capture key components of the physics at play. Part of the problem is the complexity of the phonon relaxation processes and the uncertainty underlying some of the key associated parameters; further, the model relies on a poorly defined "cavity", created by the emitter itself as it breaks the waveguide symmetry[20]. Additional work will, therefore, be mandatory to attain quantitative agreement with experiment.

**VII. Extension to polarized waveguide modes**

The description above ignores the polarized nature of the waveguide modes and thus cannot account for the observed PL polarization response. To suitably extend our description, we first note that NV relaxation from state $|E\rangle$ — in the case of NV$^-$, a six-state manifold corresponding to an orbital doublet and a spin triplet — will result in linear- or circularly-polarized photons depending on the selection rule at play[26]. At room temperature, however, dynamic Jahn-Teller averaging should yield unpolarized ensemble emission, consistent with the observations of Fig. S6a. We capitalize on this process to lay out a simplified framework where PL emission in the absence of a coupled photonic structure is assumed to be right- or left-circularly polarized with equal probabilities. In the presence of the nanodiamond and supporting AFM tip, we extend Eq. (S1) to include right- and left-circularly polarized modes, and rewrite the system Hamiltonian as

$$H = \omega_E \sigma_+ \sigma_- + \omega_W a_R^\dagger a_R + \omega_W a_L^\dagger a_L + g_R(a_R^\dagger \sigma_- + a_R \sigma_+) + g_L(a_L^\dagger \sigma_- + a_L \sigma_+), \quad (S5)$$

where the notation is the same as before except that $a_R$ ($a_L$) denotes the annihilation operator of a right (left) circularly polarized photon, and $g_R$ ($g_L$) is the emitter's coupling to the right (left) circularly polarized waveguide mode (in general, a function of the emitter's position **r**′). In the Markov approximation, we write the system's master equation as

$$\frac{d\rho}{dt} = i[\rho, H] + \kappa \mathcal{L}_R + \kappa \mathcal{L}_L + \gamma \mathcal{L}_E + \gamma^* \mathcal{L}_D. \quad (S6)$$

In the above expression, $\rho$ denotes the system's density matrix, $\kappa$ is the photon loss rate, $\gamma$ is the emitter spontaneous decay rate, and $\gamma^*$ is the emitter pure dephasing rate; the Lindbladians $\mathcal{L}$ are given by

$$\mathcal{L}_M = a_M^\dagger \rho a_M - \frac{1}{2}(a_M^\dagger a_M \rho + \rho a_M^\dagger a_M); \quad M: R, L \quad (S7)$$

$$\mathcal{L}_E = \sigma_+ \rho \sigma_- - \frac{1}{2}(\sigma_+ \sigma_- \rho + \rho \sigma_+ \sigma_-). \quad (S8)$$

Using $|\downarrow\rangle = |V\rangle$ and $|\uparrow\rangle = |E\rangle$ to denote the emitter states, and $|R, L\rangle$ to indicate the number of right- and left-circularly polarized photons, we introduce the basis set $\{|\downarrow, 0, 0\rangle, |\uparrow, 0, 0\rangle, |\downarrow, 1, 0\rangle, |\uparrow, 1, 0\rangle, |\downarrow, 0, 1\rangle, |\uparrow, 0, 1\rangle, |\downarrow, 1, 1\rangle, |\uparrow, 1, 1\rangle\}$, and define the pure dephasing Lindbladian as

$$\mathcal{L}_D = -\frac{1}{2}\begin{bmatrix} 0 & 0 & 0 & 0 & 0 & 0 & 0 & 0 \\ 0 & 0 & \rho_{23} & 0 & \rho_{25} & 0 & 0 & 0 \\ 0 & \rho_{32} & 0 & 0 & 0 & 0 & 0 & 0 \\ 0 & 0 & 0 & 0 & 0 & 0 & \rho_{47} & 0 \\ 0 & \rho_{52} & 0 & 0 & 0 & 0 & 0 & 0 \\ 0 & 0 & 0 & 0 & 0 & 0 & \rho_{67} & 0 \\ 0 & 0 & 0 & \rho_{74} & 0 & \rho_{76} & 0 & 0 \\ 0 & 0 & 0 & 0 & 0 & 0 & 0 & 0 \end{bmatrix}. \quad (S9)$$

With these definitions, Eq. (S6) can be recast as a set of equations for the emitter and photon populations, namely

$$\frac{d\langle a_M^\dagger a_M \rangle}{dt} = ig_M(\langle a_M \sigma_+ \rangle - \langle a_M^\dagger \sigma_- \rangle) - k\langle a_M^\dagger a_M \rangle; \quad M: R, L \quad (S10)$$

and

$$\frac{d\langle\sigma_+\sigma_-\rangle}{dt} = -i\left[\sum_{M=R,L} g_M(\langle a_M\sigma_+\rangle - \langle a_M^\dagger\sigma_-\rangle)\right] - \gamma\langle\sigma_+\sigma_-\rangle; \tag{S11}$$

Similarly, we express coherences as

$$\frac{d\langle a_M\sigma_+\rangle}{dt} = i\delta\langle a_M\sigma_+\rangle + ig_M(\langle a_M^\dagger a_M\rangle - \langle\sigma_+\sigma_-\rangle) - \frac{\kappa+\gamma+\gamma^*}{2}\langle a_M\sigma_+\rangle; \quad M:R,L \tag{S12}$$

where we have introduced the detuning $\delta = \omega_E - \omega_W$. In the regime where the pure dephasing rate $\gamma^* \gg \gamma, \kappa$ — valid for room temperature NVs — one can adiabatically eliminate all coherences. In this limit, we can recast the set of equations controlling all populations as

$$\frac{d\langle a_M^\dagger a_M\rangle}{dt} = \Gamma_M(\langle a_M^\dagger a_M\rangle - \langle\sigma_+\sigma_-\rangle) - k\langle a_M^\dagger a_M\rangle; \quad M:R,L \tag{S13}$$

$$\frac{d\langle\sigma_+\sigma_-\rangle}{dt} = -\left[\sum_{M=R,L} \Gamma_M(\langle a_M^\dagger a_M\rangle - \langle\sigma_+\sigma_-\rangle)\right] - \gamma\langle\sigma_+\sigma_-\rangle; \tag{S14}$$

where we introduced the effective position-dependent coupling rates

$$\Gamma_M(\mathbf{r}') = \frac{4g_M^2(\mathbf{r}')}{\kappa+\gamma+\gamma^*}\frac{1}{1+\left(\frac{2\delta}{\kappa+\gamma+\gamma^*}\right)^2}; \quad M:R,L. \tag{S15}$$

These expressions can be seen as a generalization of those derived for an emitter coupled to a single cavity mode[17,18], and imply that the NV PL will experience not only a frequency change but will also show a preferential polarization defined by the dominating coupling constant at the nanodiamond site, a point we address immediately below.

### VIII. Numerical modeling of the NV dynamics

To rationalize the measured polarization patterns (Fig. 4 in the main text), we performed numerical simulations of the electromagnetic response of the topological photonic crystal. All calculations were carried out using COMSOL Multiphysics and we used a supercell comprising 15 unit cells in the direction transverse to the waveguide and up to 3 unit cells along the waveguide. The latter is necessary attain convergence in the results we obtain as we displace the emitter vertically. We focus on the counterpropagating waveguide modes near 700 nm, and numerically determine the spatial dependence of the dipole coupling constants over a section of the crystal intersecting the domain wall. Figures S12 through S14 summarize our results for the case of a circularly polarized dipole of varying orientation and height relative to the waveguide plane. In each case, we calculate the spatial dependence of $\tilde{g}_R$ ($\tilde{g}_L$), i.e., the coupling constant of an RCP (LCP) dipole to *both* counter-propagating modes in the waveguide. For comparison with our polarization-selective measurements, we also calculate the spatial map for $\tilde{g}_S = \tilde{g}_R + \tilde{g}_L$ as well as for $W = |\tilde{g}_S|\log(\tilde{g}_R/\tilde{g}_L)$, a metric introduced here to highlight the areas of the waveguide where one type of coupling dominates over the other.

Consistent with our observations, the dipole invariably couples with higher efficiency over a ~1 μm wide section around the domain boundary, although a comparison with experiment requires a careful consideration of the dipole orientation and height. For example, modeling for a dipole circularly polarized on a plane parallel to the photonic structure (Fig. S12a) shows that for heights between 50 and 150 nm above the structure (compatible with the conditions in our experiments) the dominating coupling alternates from $\tilde{g}_R$ to $\tilde{g}_L$ as one moves along the waveguide (see bottom $W$ plots in Figs. S12b through S12g). However, the overall pattern shows antisymmetry across the waveguide (best seen in the four-unit-cell plot on the right-hand side of Fig. S12h), which we do not observe experimentally.

The $W$ plots acquire approximate mirror symmetry across the waveguide when we tilt the plane of the dipole so as to include the normal to the photonic crystal surface. Figure S13 illustrates this scenario for the case where the dipole plane contains the *z*- and *y*-axes: Curiously, dipole heights of 300 and 400 nm above the structure lead to alternating fringes analogous to those observed experimentally. These distances, however, are much greater than those in our measurements suggesting the similarity is fortuitous; further, for a given type of coupling, we only find one fringe per unit cell, inconsistent with our observations. We note that the agreement does not improve if rather than this geometry we consider an alternative where the dipole plane contains the *z*- and *x*-axes (not included here for brevity).

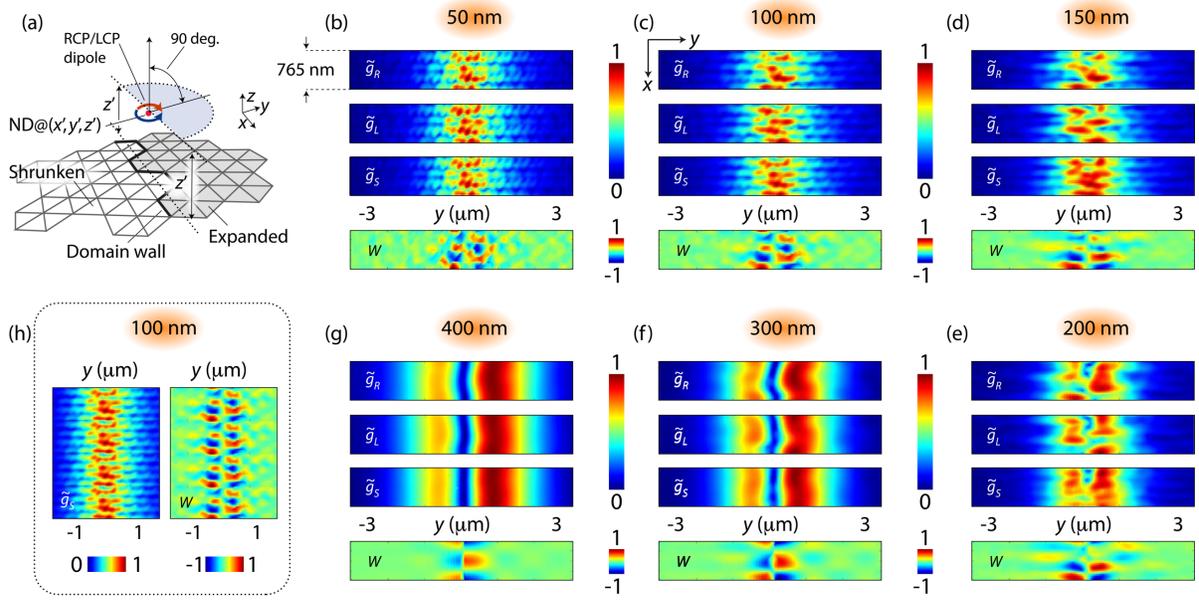

**Figure S12 | Numerically modeling the coupling of a circularly polarized dipole to the waveguide modes.** (a) We consider a circularly polarized dipole on the plane of the waveguide. The geometry is identical to that in Fig. 4 of the main text except that now we vary the dipole height $z'$. (b through g) Starting from the top, each column shows the normalized coupling constant map for a right- and left-circularly polarized dipole, respectively $\tilde{g}_R$ and $\tilde{g}_L$, as well as the sum of the two, i.e., $\tilde{g}_S = \tilde{g}_R + \tilde{g}_L$; the last rectangle displays the map for $W = |\tilde{g}_S| \log(\tilde{g}_R / \tilde{g}_L)$. The area in each rectangle corresponds to a strip of the photonic crystal across a domain wall stretching one waveguide period along $x$; we note, however, our calculations made use of a supercell comprising three identical consecutive strips, a condition required to attain good convergence. (h) Coupling constant pattern along four waveguide periods assuming 100 nm dipole height. While $\tilde{g}_S$ nearly forms a continuum along the waveguide (left image), the polarization-selective pattern implied by $W$ (right hand image) markedly differs from that observed experimentally.

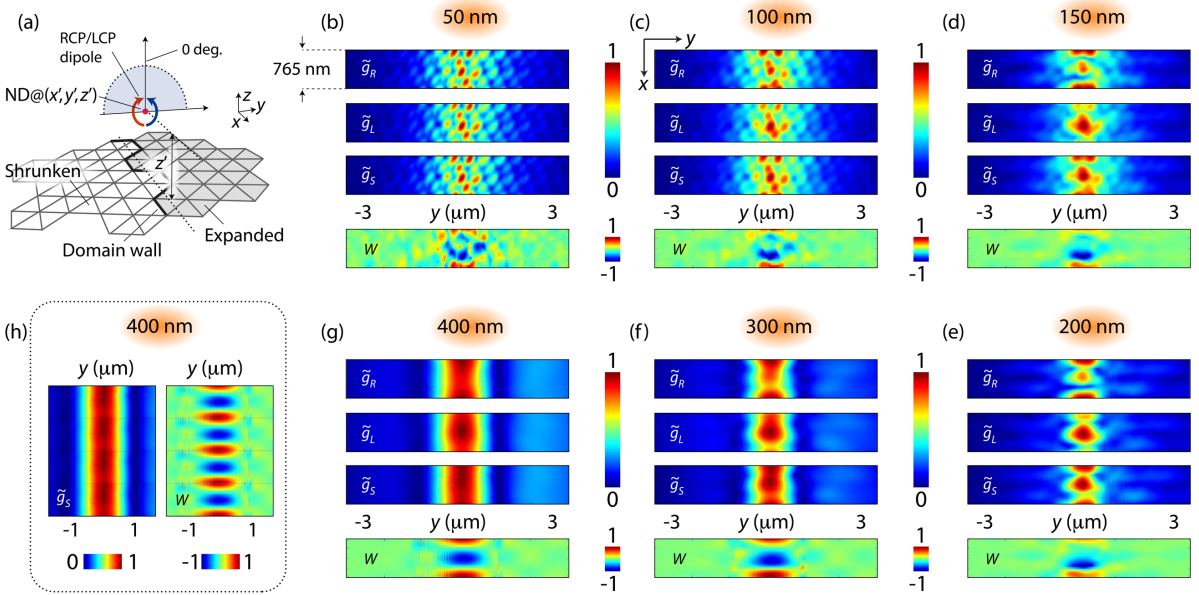

**Figure S13 | Numerically modeling the coupling of a circularly polarized dipole to the waveguide modes (continued 1).** (a) Same as in S12a but for a dipole circularly polarized on the plane containing the $z$- and the $y$-axes. (b through g) Same as in S12a through S12g but for the dipole in (a). (h) Coupling constant pattern along four waveguide periods assuming the dipole height is 400 nm. We find that $g_S$ nearly forms a continuum along the waveguide while the polarization selective map forms of a set of alternating fringes. Nonetheless, the number of fringes corresponding to one helicity (one per unit cell) is half of that observed experimentally; further, the 400 nm height required to observe such a pattern is considerably greater than in our measurements (50 to 100 nm).

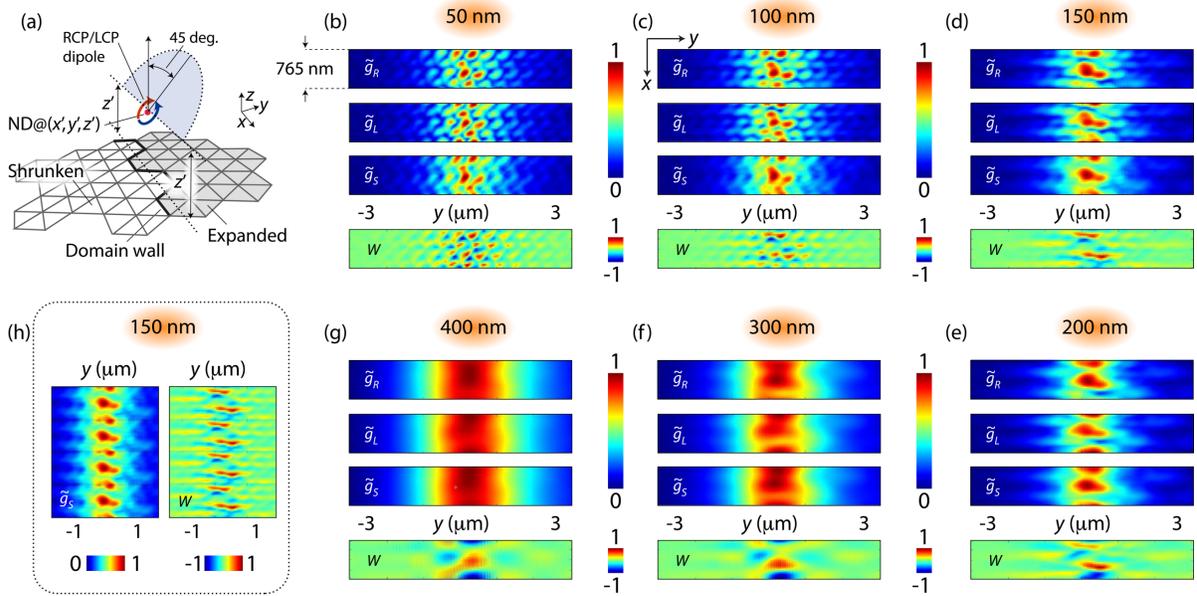

**Figure S14 | Numerically modeling the coupling of a circularly polarized dipole to the waveguide modes (continued 2).** (a) Same as in S12a but for a dipole circularly polarized on a plane forming a 45 deg. angle with the *z*-axis. (b through g) Same as in S12a through S12g but for the dipole in (a). (h) Coupling constant pattern along four waveguide periods assuming the dipole height is 100 nm. The locations where one type of coupling dominates over the other tend to align resulting in dashed fringes of alternating preferred ellipticity. Unlike in our experiments, however, the simulations lead to four fringes of one type per unit cell, double the number in the measured polarization patterns.

An interesting alternative, however, is that considered in Fig. S14. In this case, the plane of the circularly polarized dipole forms a 45 deg. angle with the *z*-axis and is oriented such that the intersection with the *xy*-plane coincides with the direction of the waveguide. For heights between 50 and 150 nm, the $W$ plots exhibit a collection of high coupling spots, gradually merging to form "quasi" fringes of alternating sign. Furthermore, for a given coupling type, we find two fringes per unit cell, as observed experimentally (consider, for example, the four-unit-cell map on the right hand side of Fig. S14h). Fringes of the same sign form along the waveguide slightly off center to generate a zig-zag; although not apparent, this feature is also present in our polarization sensitive data set, particularly in the fringe pattern along boundary $y_b^{(2)}$ in Fig. 4a of the main text.

While the agreement we attain with our observations should be deemed qualitative, several simplifications necessarily make our model a crude proxy of the physical system under consideration. One example is the use of a single effective dipole to capture the integrated response of all NVs in the nanodiamond-hosted ensemble, a concept at odds with the extended size of the particle (~75 nm in the experiments of Fig. 4) and the sensitive dependence on the dipole distance to the surface. Another important consideration is the impact of the silicon tip, ignored in our simulations but likely to alter the LDOS of the bare photonic crystal near the diamond nanoparticle.

One concrete demonstration of this latter interplay is shown in Fig. S15, which summarizes polarization-sensitive observations similar to those already discussed, but in a geometry where the AFM cantilever — and, correspondingly, the in-plane projection of the tip — aligns with the direction of the photonic waveguide (see schematics in Figs S15a and S15b). Similar to the results in Fig. 4 of the main text, PL scans at 700 nm result in periodic fringe patterns of alternating ellipticity or a trench-like structure, depending on whether we make detection sensitive to circular or linear polarization (Figs. S15c through S15e). Remarkably, however, this modified, "longitudinal" geometry yields fringes featuring substantially smaller transverse spread (of order ~0.3 nm); further, we find that the fringe centers displace laterally to form a zig-zag pattern only marginally noticeable in the polarization-selective maps of Fig. 4 in the main text. We also observe some asymmetry in the trench-like patterns produced by monitoring the NV linear polarization at 700 nm or the PL amplitude at 660 nm (Figs. S15e and S15f, respectively).

The mechanism at play is presently unclear but we hypothesize the AFM tip — which, as the photonic crystal, is made out of silicon — modifies the LDOS in a non-trivial manner, hence altering the NV response. In particular, we suspect the tapered geometry of the tip can act itself as an antenna. The NVs, therefore, effectively experience a three-dimensional photonic structure combining the topological waveguide and the AFM tip into a single whole that simultaneously imprints and guides the PL photons into the objective. Qualitatively, this picture helps us rationalize the large spectral and polarization contrast in the NV emission under fast pure dephasing, substantially greater than that reported before[6]. By the same token, the orientation of the AFM tip is expected to influence the ability to probe local mode structure and thus affect the imaging resolution that can be attained. A full modeling of this process, however, exceeds the scope of this article, and we postpone it for future work.

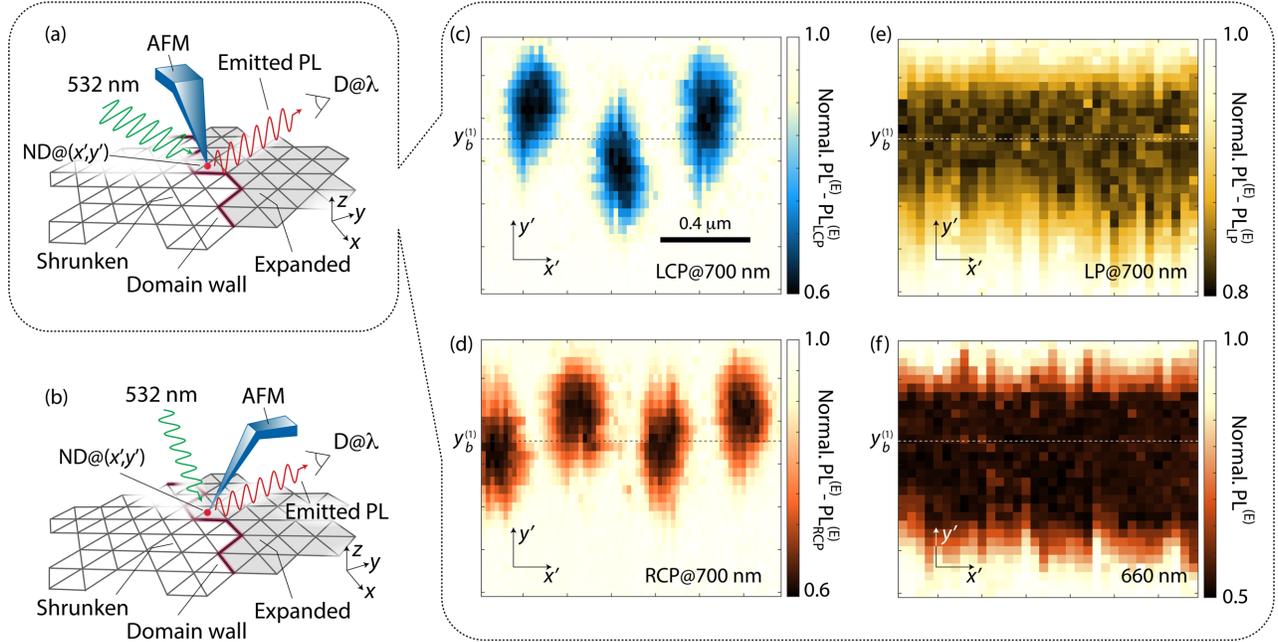

**Figure S15 | Impact of the AFM tip orientation.** (a, b) "Longitudinal" and "transverse" tip geometry — (a) and (b), respectively — featuring an AFM tip oriented parallel or perpendicular to the waveguide axis (itself running along *x*). (c, d) Respectively, LCP- and RCP-sensitive images of a section of the waveguide at $y_b^{(1)}$ with detection at 700 nm. Compared to the images in Fig. 4a of the main text, we observe a collection of fringes with the same periodicity, but a zig-zagging trend not observed before. (e) Same as before but for detection sensitive to the PL linear polarization (compare to top image in Fig. 4b of the main text). (f) Same as in (e) but for a 10 nm detection window centered at 660 nm (compare to scan images in Fig. 1c of the main text). Except for the tip relative orientation, all conditions are similar to those in Figs. 1 and 4 of the main text.

## IX. Probe robustness and data reproducibility

Attaching the nanoparticle to the AFM tip is largely the result of a process of trial and error, typically comprising multiple scans of a target nanoparticle followed by inspection of the outgoing PL. The serendipity inherent to this assembly protocol suggests some lack of robustness, hence raising questions on the long-term system response. Our experience indicates, nonetheless, that the scanning geometry is remarkably robust: As a matter of fact, *a single tip hosting the same nanodiamond* was used in all experiments reported in the main text and Figs. S1 through S13 in the Supplementary Material, thus demonstrating that the system can be operated reliably over a long time (extending over months). Further, the system is not only stable, but it also yields reproducible results: One demonstration is Fig. S9 (see Section V above), correctly replicating the fringe pattern already presented in Fig. 4 of the main text despite having been acquired several weeks later.

While a statistical analysis of the tip performance exceeds the scope of this work (and is likely to prove of limited practical value), the results in Fig. S16 provide a sense on data reproducibility. Starting in Fig. S16a, we present three different SEM images of a scanning tip: For convenience, Panel (*i*) reproduces that already introduced in Fig. S2 and corresponds to the tip/nanodiamond assembly used throughout much of this work. Panel (*ii*), on the other hand, displays a different tip after attaching a new diamond nanoparticle. As in the preceding case, the particle localizes close to the tip apex (and yields comparable performance, as we show immediately next). Admittedly, however, the assembly process largely depends on the skills of the AFM operator and is prone to failure if not carried out optimally. As an illustration, Panel (*iii*) in that same figure shows an SEM image of an AFM probe whose tip broke after pressing too hard against the substrate while trying to pick up a nanodiamond.

Importantly, the results we present do not depend on the particular nanodiamond of choice. We demonstrate this point in Fig. S16b, where we resort to the new probe in Panel (*ii*) of Fig. S16a to reconstruct a polarization-selective image. Zeroing in first on the same section of the crystal presented in Fig. 4 of the main text, we obtain nearly identical results. It is worth noting, however, that not all areas in the crystal yield the same NV response. This is shown in Fig. S16c where we instead use the tip in Panel (*i*) — already employed in Fig. 4 — to map out a different section of the photonic structure; while the fringe pattern is unmistakable, we find a fainter signature, likely the consequence of inhomogeneities introduced during the etching process. This type of variability — greater than that introduced by the particular nanodiamond/tip assembly in use — should therefore be seen as prevalent.

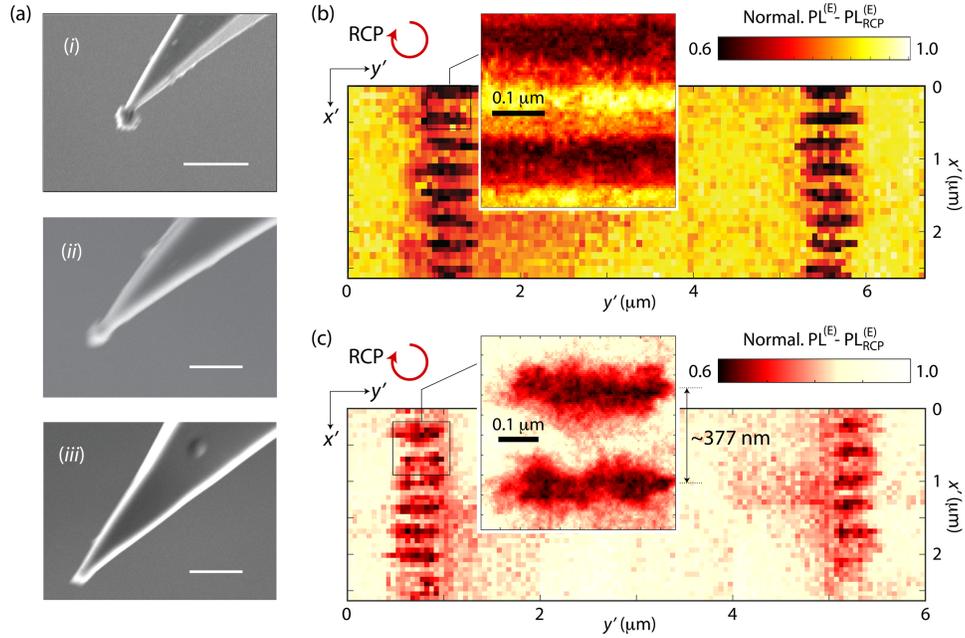

**Figure S16 | Tip assembly and measurement reproducibility.** (a) As a reference, the SEM image in (*i*) shows the tip and diamond nanoparticle used in all preceding experiments (reproduced from Fig. S2d in Section I). The image in (*ii*) corresponds to a different tip after attaching a new nanodiamond; the broken tip presented in (*iii*) shows the result after pressing too hard against the substrate and failing to attach a nanoparticle. The scale bar corresponds to 200 nm. (b) RCP-selective image of the photonic crystal using the tip/nanodiamond assembly in (*ii*); the result is comparable to that shown in Fig. 4a of the main text. (c) Same as in (b) but using the tip in (*i*) over a different section of the photonic crystal; the pattern is less resolved, likely a consequence of inhomogeneity in the structure.